%% file: catalog.tex
\documentclass[preprint,10pt]{aastex}
\pdfoutput=1
\usepackage{grffile}  

\usepackage{url}                                                                                  

\renewcommand{\S}{Section }
\newcommand{\tnm}[1]{\tablenotemark{#1}}

\newcommand{\anchorfoot}[2] {\anchor{#1}{#2}\footnote{\url{#1}}}
\newcommand{\anchorparen}[2]{\anchor{#1}{#2} (\url{#1})}

\newcommand{\todo}[1]       {[[{\bf \footnotesize #1}]]} 

\newcommand{\tbr}[1]        {#1}

\newcommand{\revise}[1]     {{\bf #1}}

\newcommand{\Spitzer} {{\em Spitzer}}
\newcommand{\Chandra} {{\em Chandra}}
\newcommand{\ACIS}    {{ACIS}}
\newcommand{\CIAO}    {{\em CIAO}}
\newcommand{\Sherpa}  {{\em Sherpa}}

\newcommand{\MARX}    {{\em MARX}}
\newcommand{\AEacro}  {{\em AE}}
\newcommand{\TARA}    {{\em TARA}}
\newcommand{\XSPEC}   {{\em XSPEC}}

\newcounter{column_number}
\newcommand{\numberthecolumn}{\colhead{(\arabic{column_number})}\stepcounter{column_number}}

\shorttitle{CCCP Catalog}
\shortauthors{Broos et al.}                                

\begin{document}

\title{A Catalog of \Chandra\ X-ray Sources in the Carina Nebula}

\author{
Patrick S. Broos\altaffilmark{1},
Leisa K. Townsley\altaffilmark{1}, 
Eric D. Feigelson\altaffilmark{1}, 
Konstantin V. Getman\altaffilmark{1}, 
Gordon P. Garmire\altaffilmark{1},
Thomas Preibisch\altaffilmark{2}, 
Nathan Smith\altaffilmark{3},
Brian L. Babler\altaffilmark{4},
Simon Hodgkin\altaffilmark{5}, 
R\'{e}my Indebetouw\altaffilmark{6},
Mike Irwin\altaffilmark{5}, 
Robert R. King\altaffilmark{7}, 
Jim Lewis\altaffilmark{5},  
Steven R. Majewski,\altaffilmark{6}, 
Mark J. McCaughrean\altaffilmark{7,8}, 
Marilyn R Meade\altaffilmark{4},
Hans Zinnecker\altaffilmark{9}
}
\email{patb@astro.psu.edu}

\altaffiltext{1}{Department of Astronomy \& Astrophysics, 525 Davey Laboratory, 
Pennsylvania State University, University Park, PA 16802, USA} 

\altaffiltext{2} {Universit\"ats-Sternwarte, Ludwig-Maximilians-Universit\"at, Scheinerstr.~1, 81679 M\"unchen, Germany}

\altaffiltext{3} {Steward Observatory, University of Arizona, 933 North Cherry Avenue, Tucson, AZ 85721, USA}

\altaffiltext{4} {University of Wisconsin Department of Astronomy, 475 N. Charter St., Madison, WI 53706, USA}

\altaffiltext{5} {Cambridge Astronomical Survey Unit, Institute of Astronomy, Madingley Road, Cambridge, CB3, UK}

\altaffiltext{6}{Department of Astronomy, University of Virginia, P. O. Box 400325, Charlottesville, VA 22904-4325, USA}

\altaffiltext{7} {Astrophysics Group, College of Engineering, Mathematics, and Physical Sciences, University of Exeter, Exeter EX4 4QL, UK}

\altaffiltext{8} {European Space Agency, Research \& Scientific Support Department, ESTEC, Postbus 299, 2200 AG Noordwijk, The Netherlands}

\altaffiltext{9} {Astrophysikalisches Institut Potsdam, An der Sternwarte 16, 14482 Potsdam, Germany}

\begin{abstract}
We present a catalog of ${\sim}$14,000 X-ray sources observed by the \ACIS\ instrument on the {\em Chandra X-ray Observatory} within a 1.42 square degree survey of the Great Nebula in Carina, known as the  \Chandra\ Carina Complex Project (CCCP).
This study appears in a \anchor{TBD}{Special Issue} of the \apjs\ devoted to the CCCP.
Here, we describe the data reduction and analysis procedures performed on the X-ray observations, including calibration and cleaning of the X-ray event data, point source detection, and source extraction.
\revise{The catalog appears to be complete across most of the field to an absorption-corrected total-band luminosity of ${\sim}10^{30.7}$ erg~s$^{-1}$ for a typical low-mass pre-main sequence star.}
Counterparts to the X-ray sources are identified in a variety of visual, near-infrared, and mid-infrared surveys.
The X-ray and infrared source properties presented here form the basis of many CCCP studies of the young stellar populations in Carina.
\end{abstract}


\section{INTRODUCTION \label{intro.sec}}

The \Chandra\ Carina Complex Project (CCCP) is a 1.42 square degree survey of the Great Nebula in Carina with the Imaging array of the Advanced CCD Imaging Spectrometer \citep[\ACIS-I,][]{Garmire03} on the {\em Chandra X-ray Observatory}. 
\citet{Townsley11a} discuss the survey motivation and design, provide an observing log for the \ACIS\ observations, and present several figures that place the footprint of the survey in an astronomical context. 
That work and the material presented here initiate a Special Issue of the \apjs\ devoted to the CCCP.  

We describe here the data reduction and analysis procedures underlying various data products that are widely used by the remaining studies in this series.
\S\ref{extraction.sec} discusses the calibration and cleaning of the X-ray event data, point source detection, and source extraction.  
Properties of the resulting X-ray point sources are presented in \S\ref{catalog.sec}. 
Visual and infrared (IR) counterparts to the X-ray sources are reported in \S\ref{counterparts.sec}, and the reliability of the counterpart identification is discussed in \S\ref{match_outcomes.sec}. 
\S\ref{quality.sec} discusses the completeness and reliability of the X-ray catalog.

\section{DATA REDUCTION AND EXTRACTION METHODS \label{extraction.sec}}

The CCCP survey consists of \dataset[ADS/Sa.CXO#DefSet/CCCP]{38 \ACIS-I observations} totaling 1.2 Ms organized into 22 pointings \citep[][\tbr{Table~1 and Figure~4}]{Townsley11a}.
\Chandra\ surveys of similar complexity include a ${\sim}$2 Ms observation of the Galactic Center region \citep{Muno09}, the CHAMP survey of 130 extragalactic fields \citep{Kim07}, the ${\sim}$1.8~Ms observation of the extragalactic C-COSMOS region \citep{Puccetti09}, and a ${\sim}$2 Ms observation of the extragalactic CDF-S field \citep{Luo08}. 

Our data analysis techniques are discussed at length by Broos et al.~(2010), 
hereafter referred to as B10; we briefly review the procedure here.
\Chandra-\ACIS\ event data are calibrated and cleaned as described in \S3 of \citet[][]{Broos10}. 
Those procedures seek to improve the accuracy of event properties (individual event positions, alignment of the \Chandra\ coordinate system to an astrometric reference, energy calibration) and seek to discard events that are likely caused by various instrumental background components.
A cosmic ray artifact known as \anchorfoot{http://cxc.harvard.edu/ciao/why/afterglow.html}{``afterglow''} creates a particularly troublesome type of background---a group of events appearing at nearly the same location on the detector in nearly consecutive CCD frames. 
Since no single currently-available method for identifying afterglow events is appropriate for both weak and bright sources, we adopt a bifurcated workflow in which heavily-cleaned data are used for source detection and lightly-cleaned data are used for source extraction \citep[see][\S3 and Figure 1]{Broos10}.

Data from two detectors lying far off-axis in the \ACIS\ S-array, available for most observations, were calibrated and cleaned as above.
However, S-array data were not used for source detection and were not routinely extracted, due to the very poor angular resolution of the \Chandra\ mirrors at large off-axis angles.
In a few cases, very bright sources produced S-array data that were very useful for studies of individual stars \citep{Townsley11a,Parkin11}.

Candidate point sources are identified in each of the CCCP pointings individually using two methods \citep[see][\S4.2]{Broos10}.
First, twelve images (energy bands 0.5--2~keV, 2--7~keV, 0.5--7~keV in combination with four pixel sizes) are searched by the standard \Chandra\ detection tool, \anchorfoot{http://asc.harvard.edu/ciao/download/doc/detect_manual}{{\it wavdetect}} \citep{Freeman02}.
Second, \revise{Lucy-Richardson image reconstruction \citep{Lucy74} is performed on overlapping images (0.5--8~keV) that tile each CCCP pointing, sized ($1.5\arcmin \times 1.5\arcmin$) so that the \Chandra\ PSF is relatively constant within each tile; peaks in the reconstructed tiles are adopted as candidate sources.
}
Candidate sources are {\em not} obtained from observations in other bands (e.g., visual or IR).

These candidate sources are extracted from all the observations in which they appear on the \ACIS\ I-array using the \anchorfoot{http://www.astro.psu.edu/xray/acis/acis_analysis.html}{{\it ACIS Extract}} (\AEacro) package \citep[see][\S5]{Broos10}.
Sixty percent of CCCP sources were observed multiple (2--7) times, as illustrated by the CCCP exposure map, which is shown in Figure~\ref{NetCounts_sample.fig} here, \tbr{Figure~4} in \citet[][]{Townsley11a}, and Figure~2 in \citet[][]{Broos10}.
The range of off-axis angles, and thus the range of extraction aperture sizes, found among the multiple observations of a single source is often small---when observations share the same pointing or when observations barely overlap---but can be large when observations moderately overlap.

The positions of source candidates are updated with \AEacro\ estimates, using a subset of each source's extractions chosen to minimize the position uncertainty \citep[see][\S6.2 and 7.1]{Broos10}.  
A source significance statistic is calculated with respect to the local background level \citep[see][\S4.3]{Broos10}, using a subset of each source's extractions chosen to maximize that significance \citep[see][\S6.2]{Broos10}.
This \revise{decision} to ignore some observations of a source effectively prevents the very poor point spread function (PSF) of an off-axis observation from spoiling the detection or position estimate of a source that was also observed with a very good PSF on-axis.
As the Carina field has bright astrophysical diffuse emission with complex structure, the use of a local rather than global background source existence criterion is important.  
Only candidates found to be significant are accepted as X-ray sources; expressed in terms of source properties reported in Table~\ref{xray_properties.tbl}, \revise{a detection requires both SrcCounts\_t $\ge 3$ {\em and} ProbNoSrc\_min $< 0.01$.}  
Iterative pruning of source candidates and re-extraction continues until no candidates are found to be insignificant \citep[see][\S4.1]{Broos10}.
\revise{Our selection of the ProbNoSrc\_min threshold is discussed further in \S\ref{spurious.sec}.}

Photometry, spectra, light curves, and a variety of apparent source properties (\S\ref{catalog.sec}) are derived for sources in the final catalog, using a subset of each source's extractions chosen to balance the conflicting goals of minimizing photometric uncertainty and of avoiding photometric bias \citep[see][\S6.2]{Broos10}.

The CCCP survey region includes several rich young stellar clusters, and some sparse but compact stellar groups, where source crowding can be important.  In such cases, the source detection is largely based on the maximum likelihood image reconstruction where the blurring effects of the PSF are reduced.  Photons are then extracted in apertures based on the local PSF shape but reduced in size to avoid overlap with adjacent source extraction regions.  While the procedures are well-established in \citet{Broos10} with little subjective judgment involved, it is important to recognize that the source identifications and extractions in very crowded regions represent only one plausible interpretation of the data.


The \Chandra\ data analysis system, \CIAO, \citep{Fruscione06}, the {\it SAOImage ds9} visualization tool \citep{Joye03}, and the \anchorfoot{http://www.ittvis.com/idl}{{\it Interactive Data Language}} (IDL) are used throughout our data analysis workflow, from data preparation through science analysis.
The CCCP data were reduced with \CIAO\ version 4.0.2, using CALDB version 3.4.2 for event processing and CALDB version 4.1.1 for construction of calibration data products.
Models of the \Chandra-\ACIS\ PSF were constructed \citep[see][Appendix~C]{Broos10} using version 4.3 of the \anchorfoot{http://space.mit.edu/cxc/marx/}{\MARX\ mirror simulator}.

\section{APPARENT X-RAY POINT SOURCE PROPERTIES \label{catalog.sec}}

A total of 14,369\footnote
{
The catalog used in other CCCP studies contains 14,368 sources; one additional source (the last reported in Table~\ref{xray_properties.tbl}) was identified too late in the data analysis process to be included in those studies.
}
point sources were identified and extracted.
The CCCP observations are not appropriate for studying the luminous blue variable $\eta$~Carinae, and it is omitted from our catalog even though it is a strong X-ray source; \citet{Townsley11a} provide references to more suitable observations of this remarkable star.  
The CCCP catalog is depicted in \citet[][Figure~3 and 4]{Townsley11a}, and in \citet[][\tbr{Figure~4}]{Broos11}.
No single figure can adequately represent such a wide (1.4 square degrees) \Chandra\ field since sources as close as $<$1\arcsec\ separation can be detected.
However, the electronic versions of those figures can reveal, when zoomed, detail not visible in most printed versions.

Estimates for many apparent (not corrected for absorption) properties of the X-ray sources \citep[see][\S7]{Broos10} are provided in a table published electronically and available at Vizier \citep{Ochsenbein00}.
Since the table has many columns, a stub cannot be conveniently shown in print.
Instead, column names that are appropriate for Vizier and column descriptions are listed in Table~\ref{xray_properties.tbl}.
The suffixes ``\_t'', ``\_s'', and ``\_h'' on names of photometric quantities designate the {\em total} (0.5--8~keV), {\em soft} (0.5--2~keV), and {\em hard} (2--8~keV) energy bands. 
The SrcCounts and NetCounts quantities characterize the extraction; correction for finite extraction apertures is applied to the ancillary reference file (ARF) calibration products \citep[see][\S5.3]{Broos10}.
Thus, the lowest-level calibrated photometric quantity that can be used to compare sources would be apparent {\em photon} flux \citep[see][\S7.4]{Broos10}
\begin{equation}
F_{\rm photon} \doteq {\rm NetCounts} / {\rm MeanEffectiveArea} / {\rm ExposureTimeNominal}  \label{photon_flux.eqn}
\end{equation}
which has units of photon~cm$^{-2}$~s$^{-1}$.
Rough, model-independent estimates for apparent {\em energy} flux are provided as
\begin{equation}
  {\rm EnergyFlux} \doteq 1.602 \times 10^{-9} \; {\rm MedianEnergy} \times F_{\rm photon} \label{EnergyFlux.eqn}
\end{equation}
in units of erg~cm$^{-2}$~s$^{-1}$, where the constant $1.602 \times 10^{-9}$ arises from the conversion between keV and erg \citep{Getman10}.
Table notes provide additional information regarding the definition of source properties.
Model-dependent estimates for absorption and intrinsic energy flux are provided by \citet[][\tbr{Table~8}]{Broos11} for sources that are likely to be low-mass young stars in the Carina complex.

\input{xray_column_labels}


High-quality spectra are available for relatively few CCCP sources.
Table~\ref{netcounts_breakdown.tbl} shows the number of CCCP sources that have spectra with various levels of quality (characterized by ranges of net counts in the total band, NetCounts\_t); the tallies are reported separately for a set of very significant ``primary'' sources, defined as having a low probability of being a spurious source (ProbNoSrc\_min in Table~\ref{xray_properties.tbl} less than 0.003) and the complementary set of less significant ``tentative'' sources (\revise{0.003 $<$ ProbNoSrc\_min $<$ 0.01}).

\begin{deluxetable}{rrr}
\tablecaption{Quality of Spectra for CCCP Sources \label{netcounts_breakdown.tbl}
}
\tablewidth{0pt}
\tabletypesize{\footnotesize}

\tablehead{
\colhead{NetCounts\_t} & \multicolumn{2}{c}{Number of sources}   \\
                       & \colhead{Primary} & \colhead{Tentative} \\
\colhead{(count)}      &  \\
\numberthecolumn & \numberthecolumn & \numberthecolumn 
\setcounter{column_number}{1}
}
\startdata
    $<$10       & 6852 & 1424 \\
 10--\phn\phn50 & 4934 &   70 \\
 50--\phn100    &  677 &    0 \\
100--\phn500    &  371 &    0 \\
500--1000       &   24 &    0 \\
  $>$1000       &   16 &    0 
\enddata
\end{deluxetable}

%

\begin{figure}[htb]
\centering
\includegraphics[width=1.0\textwidth]{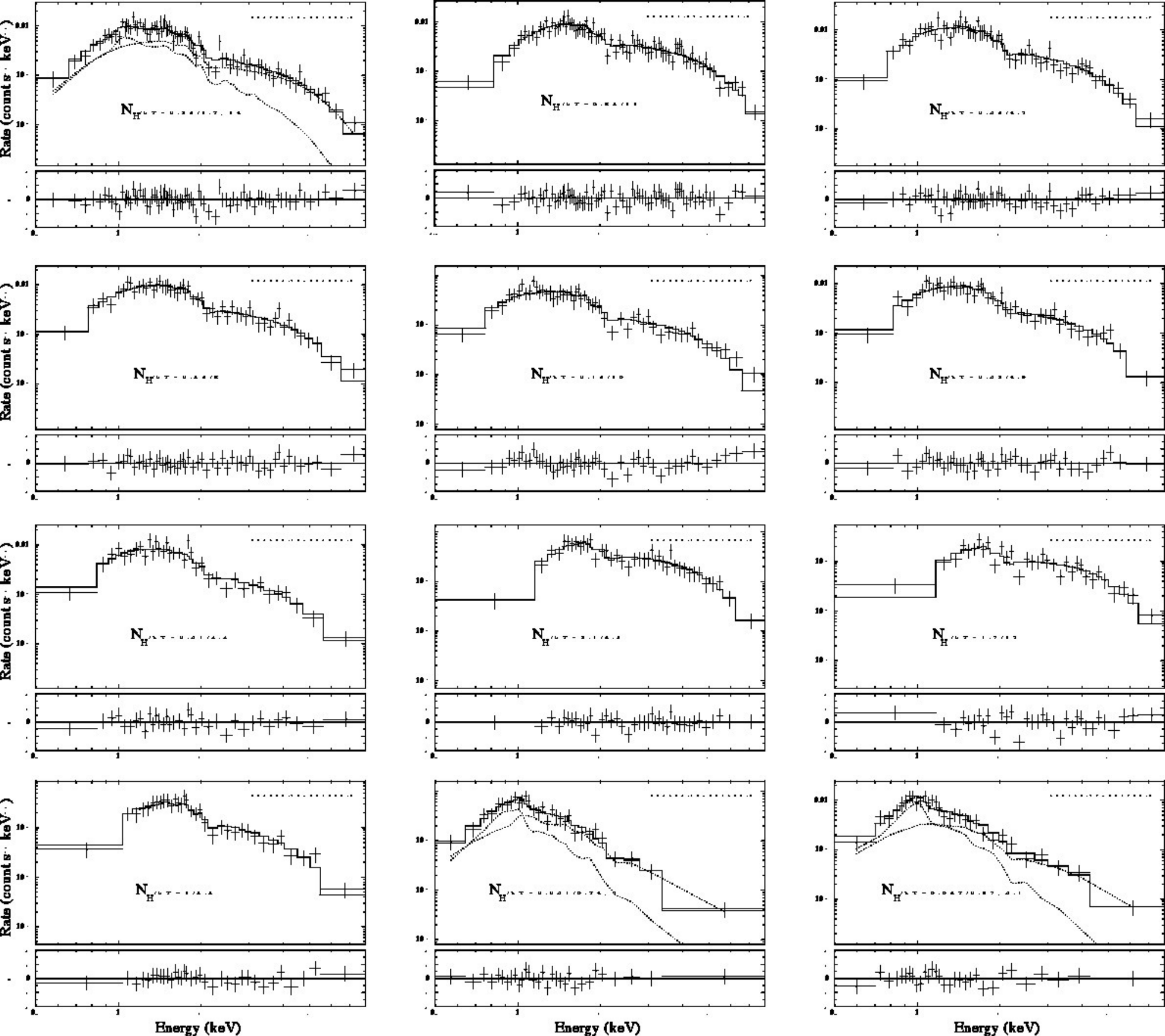}
\caption{High-quality spectra for sources from Table~\ref{high-quality-spectra.tbl} expected to be pre-main sequence stars.
The spectra are modeled as one-temperature or two-temperature thermal plasmas with abundances frozen at the values adopted by the XEST study \citep{Gudel07}, relative to  \citet{Anders89}, scaled to \citet{Wilms00}, using the {\em tbabs} absorption code.
The model is implemented in \XSPEC\ as \mbox{\texttt{tbabs*(apec)}} or \mbox{\texttt{tbabs*(apec+apec)}}. 
Column density and plasma temperatures are shown on each panel in units of $10^{22}$ cm$^{-2}$ and keV.
\label{high-quality-spectra.fig}}
\end{figure}

The 40 sources with the highest-quality spectra (more than 500 net counts) are listed in Table~\ref{high-quality-spectra.tbl}.
Most are known or candidate OB stars that are discussed in other CCCP studies \citep{Naze11,Povich11a,Parkin11}.
The three known Wolf-Rayet stars in Carina (WR~25, WR~24, WR~22) are discussed by \citet{Townsley11a}.
The neutron star 104608.71-594306.4 (E1\_85) is discussed by \citet{Hamaguchi09}.
The 12 remaining sources in the table exhibit the time variability and high-temperature spectra (shown in Figure~\ref{high-quality-spectra.fig}) characteristic of pre-main sequence stars during flares attributed to magnetic connection events; their inferred time-averaged intrinsic luminosities range from ${\sim}10^{31.6}$~erg~s$^{-1}$ to ${\sim}10^{32.4}$~erg~s$^{-1}$ in the total-band (0.5--8~keV).

\begin{deluxetable}{clrllr}
\tablecaption{CCCP sources with $>$500 net counts in I-array extractions \label{high-quality-spectra.tbl}
}
\tablewidth{0pt}
\tabletypesize{\tiny}

\tablehead{
\colhead{CXOGNC~J\tnm{b}} & \colhead{Label\tnm{a}} & \colhead{NetCounts\_t} & \colhead{Name} & \colhead{SpType} & \colhead{CCCP citation} \\
                          &                        & \colhead{(count)}      &  \\
\numberthecolumn & \numberthecolumn & \numberthecolumn &\numberthecolumn & \numberthecolumn & \numberthecolumn 
\setcounter{column_number}{1}
}
\startdata
104410.39-594311.1 & CTr16\_280  & 24609 &  WR 25 (\object{HD 93162})       & WN6h + OB?                & \citep{Townsley11a}\\   
104544.13-592428.1 & C4\_2144    & 19161 &  \object{HD 93403}               & O5.5(f)                   & \citep{Naze11}   \\
104445.04-593354.6 & CTr14\_3535 &  9274 &  \object{HD 93250}               & O3.5V((f+))               & \citep{Naze11}   \\
104422.91-595935.9 & C2\_1111    &  4169 &  QZ Car (\object{HD 93206})      & O9.5I                     & \citep{Parkin11} \\
104357.47-593251.3 & CTr14\_1925 &  2977 &  \object{HD 93129A}              & O2If*                     & \citep{Naze11}   \\
104352.25-600704.0 & C2\_547     &  2564 &  WR 24 (\object{HD 93131})       & WN6-A                     & \citep{Townsley11a}\\
104508.23-594607.0 & CTr16\_3102 &  1909 &  \object{Cl* Trumpler 16 MJ 496} & O8.5V                     & \citep{Naze11}   \\
103909.94-594714.5 & SB2\_6      &  1864 &                                  & candidate OB star         & \citep{Povich11a}\\
104505.90-594006.0 & CTr16\_2805 &  1653 &  \object{HD 303308}              & O4V((f+))                 & \citep{Naze11}   \\
104752.54-600215.2 & C5\_1260    &  1573 &                                  &                           & Figure~\ref{high-quality-spectra.fig} \\
104456.27-593830.4 & CTr16\_1450 &  1458 &                                  &                           & Figure~\ref{high-quality-spectra.fig} \\
104433.74-594415.4 & CTr16\_749  &  1408 &  \object{HD 93205}               & O3.5V((f+)) + O8V         & \citep{Naze11}   \\
104246.53-601207.0 & SP2\_103    &  1403 &                                  & candidate OB star         & \citep{Povich11a}\\
104402.75-593946.0 & CTr16\_172  &  1278 &                                  & candidate OB star         & \citep{Povich11a}\\
104154.91-594123.6 & C1\_99      &  1249 &                                  & candidate OB star         & \citep{Povich11a}\\
104815.18-594319.7 & E4\_67      &  1084 &                                  &                           & Figure~\ref{high-quality-spectra.fig} \\
104243.76-593954.2 & C1\_549     &   976 &                                  &                           & Figure~\ref{high-quality-spectra.fig} \\
104441.80-594656.4 & CTr16\_1028 &   976 &  \object{Cl Trumpler 16 100}     & O5.5V                     & \citep{Naze11}   \\
104055.32-594239.7 & SB1\_81     &   955 &                                  &                           & Figure~\ref{high-quality-spectra.fig} \\
104417.54-595350.5 & C2\_995     &   822 &                                  &                           & Figure~\ref{high-quality-spectra.fig} \\
104436.23-600529.0 & C2\_1238    &   820 &  \object{HD 93222}               & O8III((f))                & \citep{Naze11}   \\
104356.16-593250.9 & CTr14\_1731 &   716 &                                  &                           & Figure~\ref{high-quality-spectra.fig} \\
104357.65-593253.7 & CTr14\_1963 &   685 &  \object{HD 93129B}              & O3.5V((f+))               & \citep{Naze11}   \\
104609.90-602635.5 & SP4\_114    &   680 &                                  &                           & Figure~\ref{high-quality-spectra.fig} \\
104837.74-601325.7 & SP3\_222    &   668 &  LS 1932 (\object{HD 93843})     & O5V                       & \citep{Naze11}   \\
104117.50-594037.0 & SB1\_176    &   664 &  WR 22   (\object{HD 92740})     & WN7-A                     & \citep{Townsley11a}\\
104516.52-594337.1 & CTr16\_3341 &   624 &  \object{Cl Trumpler 16 112}     & O5.5/6V((f+?p)) + B2III/V & \citep{Naze11}   \\
104512.86-594942.3 & CTr16\_3272 &   619 &                                  &                           & Figure~\ref{high-quality-spectra.fig} \\
104455.79-595826.9 & C3\_210     &   618 &                                  &                           & Figure~\ref{high-quality-spectra.fig} \\
104343.29-602309.5 & SP2\_528    &   600 &                                  &                           & Figure~\ref{high-quality-spectra.fig} \\
104457.51-595429.5 & C3\_225     &   574 &                                  & candidate OB star         & \citep{Povich11a}\\
104605.70-595049.5 & E1\_74      &   544 &  LS 1886 (\object{HD 305525})    & O4V                       & \citep{Naze11}   \\
104615.77-592728.0 & C6\_759     &   529 &                                  &                           & Figure~\ref{high-quality-spectra.fig} \\
104408.84-593434.4 & CTr14\_2688 &   522 &  \object{HD 93161a}              & O8V + O9V                 & \citep{Naze11}   \\
104608.71-594306.4 & E1\_85      &   522 &                                  &                           & \citep{Hamaguchi09}\\
104354.40-593257.4 & CTr14\_1506 &   519 &  \object{HD 93128}               & O3.5V((f+))               & \citep{Naze11}   \\
104220.83-590908.6 & C7\_58      &   511 &                                  & candidate OB star         & \citep{Povich11a}\\
104712.63-600550.8 & C5\_846     &   502 &  \object{HD 93632}               & O5III(f)                  & \citep{Naze11}   \\
104557.13-595643.1 & C3\_1477    &   500 &  \object{Hen 3- 485}             & em                        & \citep{Gagne11}
\enddata
\tablenotetext{a}{Source labels identify a CCCP pointing \citep[][Table~1]{Townsley11a}; they do not convey membership in astrophysical clusters.}

\tablenotetext{b}{IAU source name; prefix is CXOGNC~J ({\em Chandra X-ray Observatory} Great Nebula in Carina)}
\end{deluxetable}

Photon-counting detectors, such as \ACIS, can suffer from a non-linearity known as \anchorfoot{http://cxc.harvard.edu/ciao/why/pileup_intro.html}{{\em photon pile-up}} when multiple X-ray photons arrive with a separation in time and space that is too small to allow each to be detected as a separate X-ray event.
Pile-up effects include photometric and spectroscopic miscalibration of the observation. 
Total-band photometry is underestimated, because multiple photons interact to produce only one or zero events.
The shape of the detected X-ray spectrum is hardened, because the energy assigned to a piled event will represent that from multiple photons. 

\AEacro\ screens all point source extractions for photon pile-up by estimating the observed count rate falling on an event detection cell of size 3x3 \ACIS\ pixels, centered on the source position.
Pile-up was confirmed and quantified for the at-risk extractions so identified, using an experimental Monte Carlo forward-modeling approach that reconstructs a pile-up free \ACIS\ spectrum from a piled \ACIS\ observation.
Appendix~\ref{pile-up_recon.sec} describes this method and illustrates its application to CCCP sources.
Table~\ref{pile-up_risk.tbl} lists extractions of CCCP sources---all known massive stars---that were confirmed to have significant pile-up using this method.\footnote{\revise
{Although source 104544.13-592428.1=C4\_2144=HD~93403 had a very high event rate (0.17 count~s$^{-1}$), reconstruction of the spectrum indicated very little pile-up (a value of 1.02 in column 7 of Table~\ref{xray_properties.tbl}) because the source was observed very far off-axis (8\arcmin) and thus had only a modest event rate per detection cell.
}}
Column (6) reports the ratio of the pile-up free to observed (piled) count rate in the total (0.5--8~keV) energy band.\footnote
{
We choose not to use the terms ``pile-up fraction'' or ``pile-up percentage'' because the \ACIS\ community has several conflicting definitions for those terms; see \S1.2 in \anchorparen{http://cxc.harvard.edu/ciao/download/doc/pileup_abc.pdf}{The \Chandra\ ABC Guide to Pileup}.
}
For these piled sources, Table~\ref{xray_properties.tbl} reports total-band  photometry (columns SrcCounts\_t and NetCounts\_t) as observed, (without correction) and all other quantities that are affected by pile-up are omitted.

\begin{deluxetable}{lllccccc}
\tablecaption{Sources exhibiting photon pile-up \label{pile-up_risk.tbl}
}
\tablewidth{0pt}
\tabletypesize{\footnotesize}

\tablehead{
\colhead{CXOGNC J} & \colhead{Label} & \colhead{Name} & \colhead{Observation} & \colhead{$\theta$} & \colhead{PsfFraction}& \colhead{Correction}  \\  
                                                                           &&&& \colhead{(\arcmin)} \\
\numberthecolumn & \numberthecolumn & \numberthecolumn & \numberthecolumn & \numberthecolumn & \numberthecolumn & \numberthecolumn  
\setcounter{column_number}{1}
}
\startdata
 104357.47-593251.3 & CTr14\_1925 & HD 93129A         & 4495 & 0.3 & 0.61  & 1.27 \\  
 104410.39-594311.1 & CTr16\_280  & WR 25 (HD 93162)  & 6402 & 4.8 & 0.90  & 1.25 \\  
 104433.74-594415.4 & CTr16\_749  & HD 93205          & 6402 & 1.8 & 0.50  & 1.16 \\  
 104422.91-595935.9 & C2\_1111    & QZ Car (HD 93206) & 9482 & 3.4 & 0.90  & 1.11 \\  
 104445.04-593354.6 & CTr14\_3535 & HD 93250          & 4495 & 6.3 & 0.89  & 1.05 \\  
\enddata
\tablecomments{
Col.\ (1): IAU designation ({\em Name} in Table~\ref{xray_properties.tbl})
\\Col.\ (2): Source name used within the CCCP project ({\em Label} in Table~\ref{xray_properties.tbl})
\\Col.\ (3): Object common name 
\\Col.\ (4): Observation ID 
\\Col.\ (5): Off-axis angle ({\em Theta} in Table~\ref{xray_properties.tbl})
\\Col.\ (6): Fraction of the PSF (at 1.497 keV) enclosed within the extraction region ({\em PsfFraction} in Table~\ref{xray_properties.tbl}). A reduced PSF fraction (significantly below 90\%)  indicates that the source is in a crowded region. 
\\Col.\ (7): Estimated ratio of pile-up free to observed (piled) count rates in the 0.5--8~keV energy band.
}
\end{deluxetable}

\section{COUNTERPARTS TO X-RAY POINT SOURCES \label{counterparts.sec}}

Many studies in this Special Issue rely on counterparts to CCCP sources observed in other bands.
\citet{Broos11} assign to each source a probability of Carina membership based in part on near- and mid-infrared photometry and on visual spectroscopy.
CCCP studies of massive stars rely on visual spectroscopy for conclusive identification \citep{Gagne11,Naze11,Parkin11} and on visual or infrared photometry to identify candidates \citep{Evans11,Povich11a}.
Both X-ray and infrared data are essential to studies of pre-main sequence stars \citep{Preibisch11,Wang11,Wolk11}.

Table~\ref{cat_sum1.tbl} lists 19 counterpart catalogs that we have correlated with the CCCP source list.  
The Skiff, KR, PPMXL, UCAC3 (and its associated BSS), 2MASS, and \Spitzer\ catalogs cover the entire CCCP field.  
The HAWK-I, SOFI, NACO, and Sana near-infrared catalogs cover portions of the field including the richest clusters.
The CMD, DETWC, MDW, MJ, CP, and DAY visual photometry catalogs survey small fields on the major clusters.
The X-ray study of Tr~16 by \citet{Albacete08} is discussed by \citet{Wolk11}.

\begin{deluxetable}{lllrrrrr}
\tablecaption{Summary of catalogs correlated with {\it Chandra} Carina sources \label{cat_sum1.tbl}
}
\tablewidth{0pt}
\tabletypesize{\scriptsize}

\tablehead{
\colhead{Catalog} & \colhead{Scope} & \colhead{Reference} & \colhead{$N_{cat}$} & \colhead{$N_{CCCP}$} & \colhead{$\Delta$R.A.} & \colhead{$\Delta$Dec} & \colhead{$r_{\rm median}$}  \\
&&&&& \colhead{(\arcsec)} & \colhead{(\arcsec)}  & \colhead{(\arcsec)} \\                                                                    
\numberthecolumn & \numberthecolumn & \numberthecolumn & \numberthecolumn & \numberthecolumn & \numberthecolumn & \numberthecolumn  & \numberthecolumn  
\setcounter{column_number}{1}
}
\startdata 
Skiff      & Visual spectral types                   & \citet{Skiff09}      &    271  &  130  &  0.00 &  0.01 & 0.19\\
KR         & Visual photometry                       & \citet{Kharchenko09} &    363  &  114  &  0.05 &  0.02 & 0.22\\
PPMXL      & CCD proper motions (PMs)                & \citet{Roeser10}     &  30286  & 1006  &  0.00 &  0.00 & 0.41\\ 
UCAC3      & CCD PMs                                 & \citet{Zacharias10}  &  23557  & 1471  &  0.00 &  0.00 & 0.33\\
BSS        & Bright star PMs                         & \citet{Urban04}      &     47  &   18  &  0.00 &  0.00 & 0.16\\
CMD        & Photographic PMs, Tr~14, Tr~16, Cr~232  & \citet{Cudworth93}   &    577   &  141  &  0.62 &  0.59 & 0.33\\
DETWC      & Visual photometry, Tr~14 \& 16          & \citet{DeGioia01}    &    853  &  237  &  0.53 &  0.16 & 0.27\\
MDW        & Visual spectral types, Cr~228           & \citet{Massey01}     &     73\tnm{a}   
                                                                                       &   35  &\tnm{a}&\tnm{a}& 0.19\\
MJ         & Visual photometry, Tr~14 \& 16          & \citet{Massey93}     &    768  &  178  &  0.43 &  0.27 & 0.41\\
CP         & High-mass photometry, Cr 228            & \citet{Carraro01}    &   1112  &   27  & -0.64 &  0.11 & 0.68\\
DAY        & Low-mass photometry, Cr 228             & \citet{Delgado07}    &    152  &   17  & -0.05 & -0.03 & 0.81\\
HAWK-I     & Deep near-infrared photometry           & \citet{HAWKI09}      & 595148  & 6583  & -0.01 & -0.02 & 0.31\\
2MASS      & Shallow near-infrared photometry        & \citet{Cutri03}      & 130164  & 6194  &  0.00 &  0.00 & 0.37\\
SOFI       & Deep near-infrared photometry, Tr~14    & \citet{Ascenso07}    &   4739  &  849  &  0.02 &  0.00 & 0.18\\
NACO       & Deep near-infrared photometry, Tr~14    & \citet{Ascenso07}    &    178  &   54  & -0.01 & -0.01 & 0.17\\
Sana       & Deep near-infrared photometry, Tr~14    & \citet{Sana10}       &   1955  &  371  & -0.06 &  0.04 & 0.16\\
SpVela     & Mid-infrared photometry (\Spitzer)      & \citet{Povich11b}    & 130871  & 6543  &  0.00 &  0.00 & 0.38\\
SpSmith    & Mid-infrared photometry (\Spitzer)      & \citet{Smith10}      &  97901  & 3811  &  0.00 &  0.00 & 0.44\\
AC         & \ACIS\ observation of Tr~16             & \citet{Albacete08}   &   1035  &  935  &  0.19 & -0.01 & 0.18
\enddata

\tablenotetext{a}{The reported positions for two of the 75 MDW sources do not appear to coincide with any 2MASS source.  
We ignored those sources, and adopted 2MASS positions for the 73 remaining.}

\end{deluxetable}

The CCCP catalog was matched to counterpart catalogs one at a time using a simple algorithm, described in \citet[][]{Broos10} (\S8), in which the maximum acceptable separation between an X-ray source and a counterpart is based on the individual source position errors assuming Gaussian distributions, scaled so that ${\sim}$99\% of true associations should be identified as matches.
When multiple sources in the counterpart catalog satisfy the match criterion, the closest one is adopted.
The performance of the algorithm is studied in \S\ref{match_outcomes.sec}.



\begin{figure}[htb]
\centering
\includegraphics[width=0.31\textwidth]{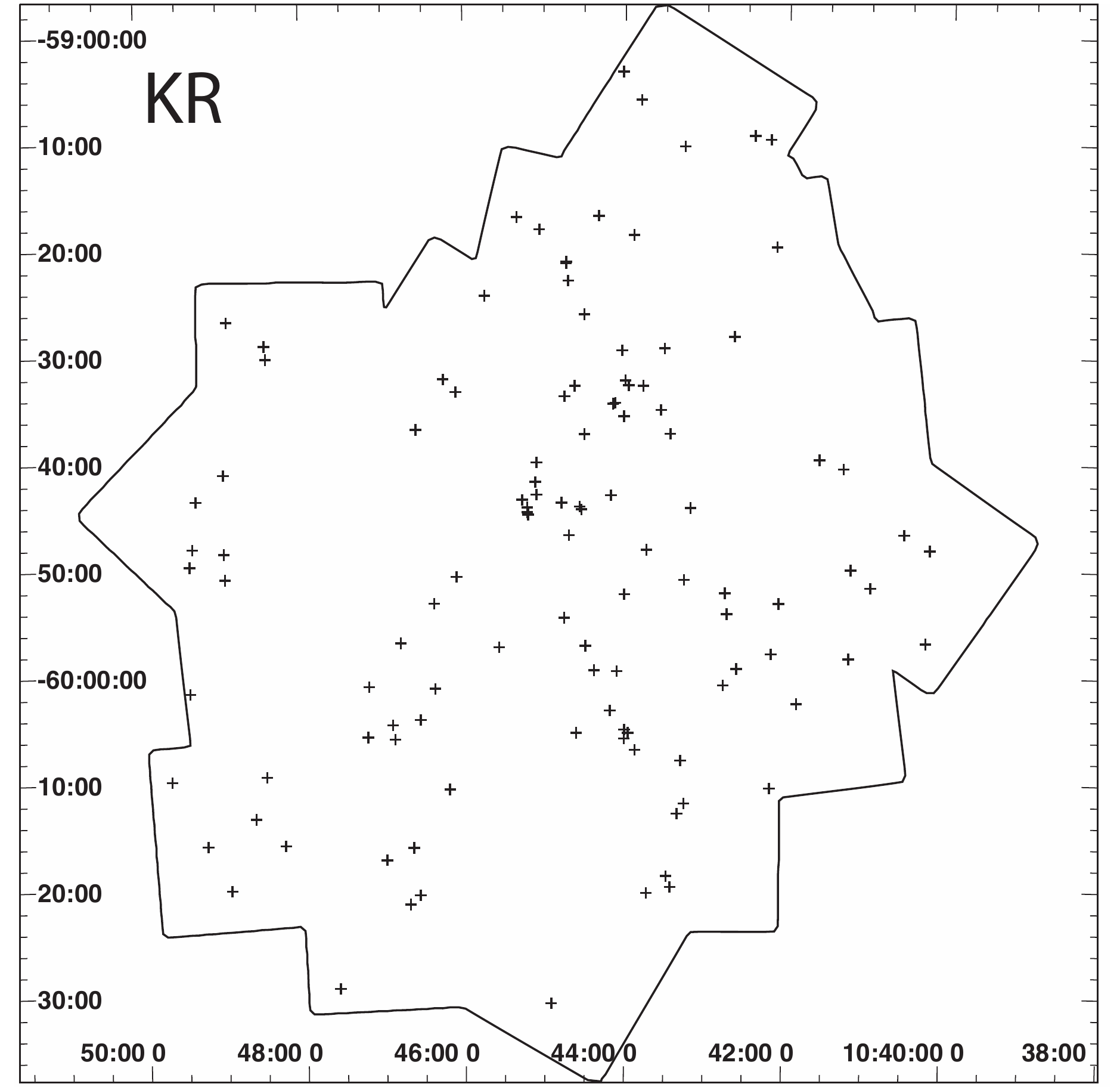}
\includegraphics[width=0.31\textwidth]{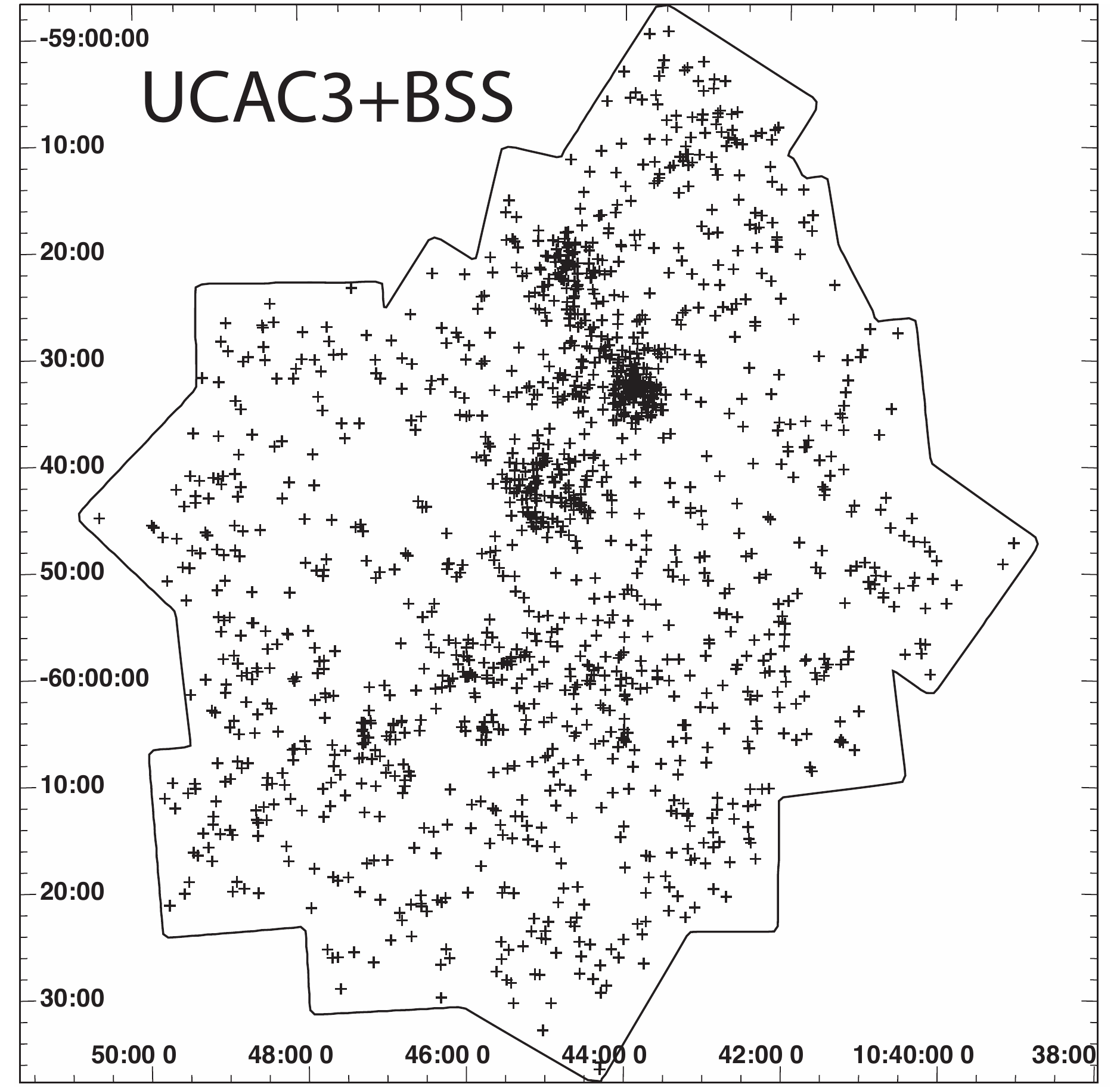}
\includegraphics[width=0.31\textwidth]{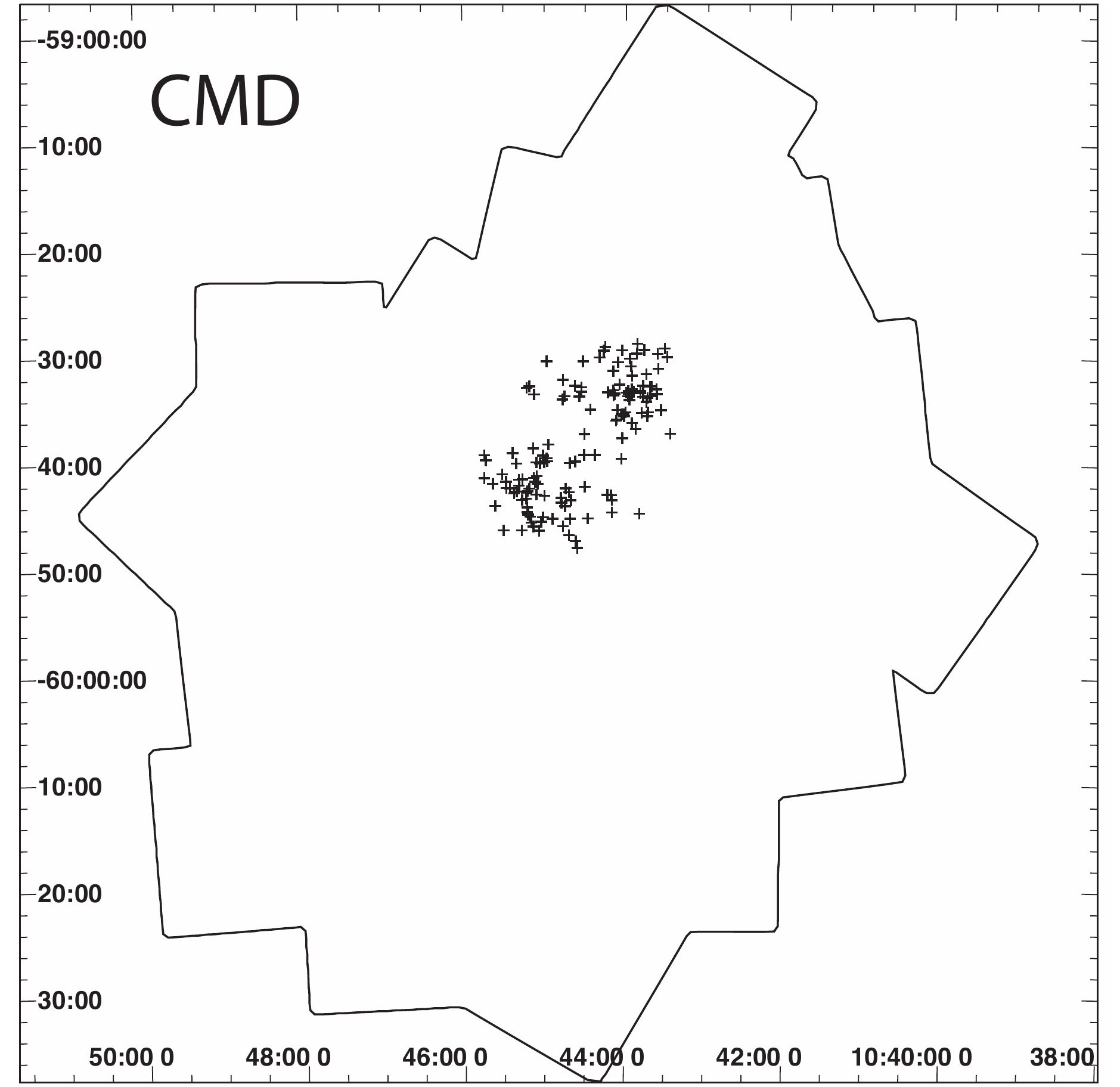}\\
 
\vspace{0.5\baselineskip} 
\includegraphics[width=0.31\textwidth]{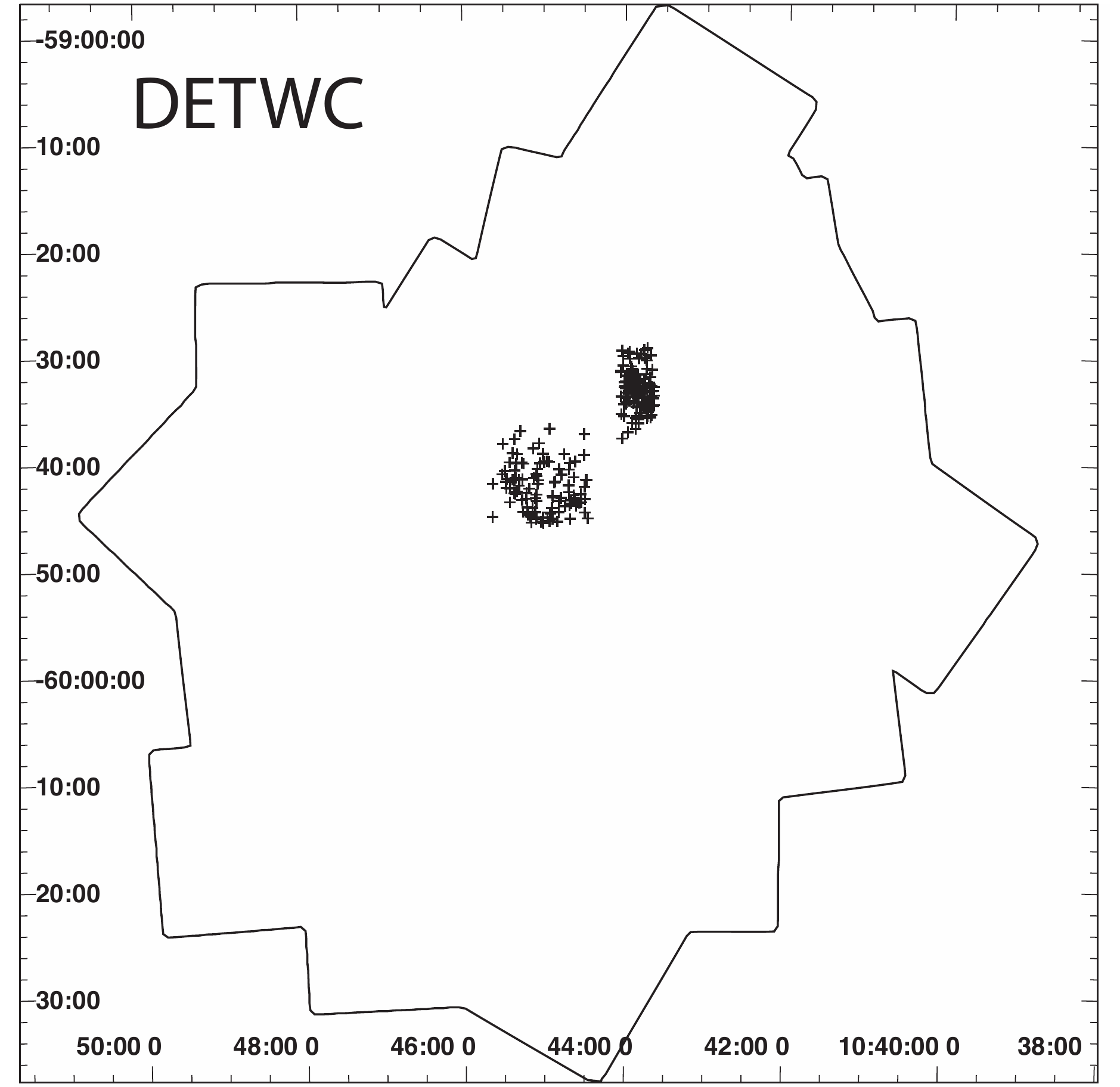}
\includegraphics[width=0.31\textwidth]{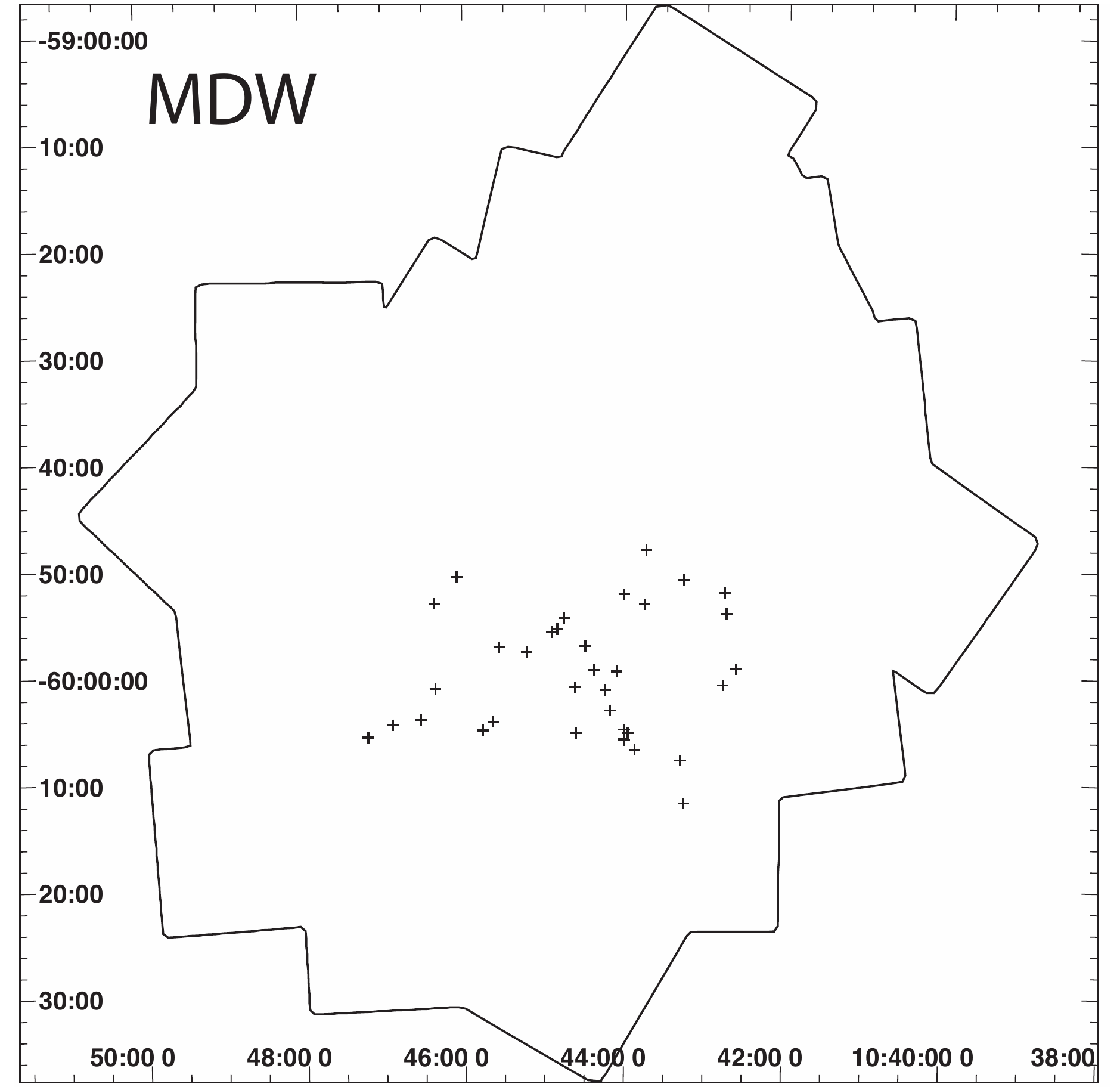}
\includegraphics[width=0.31\textwidth]{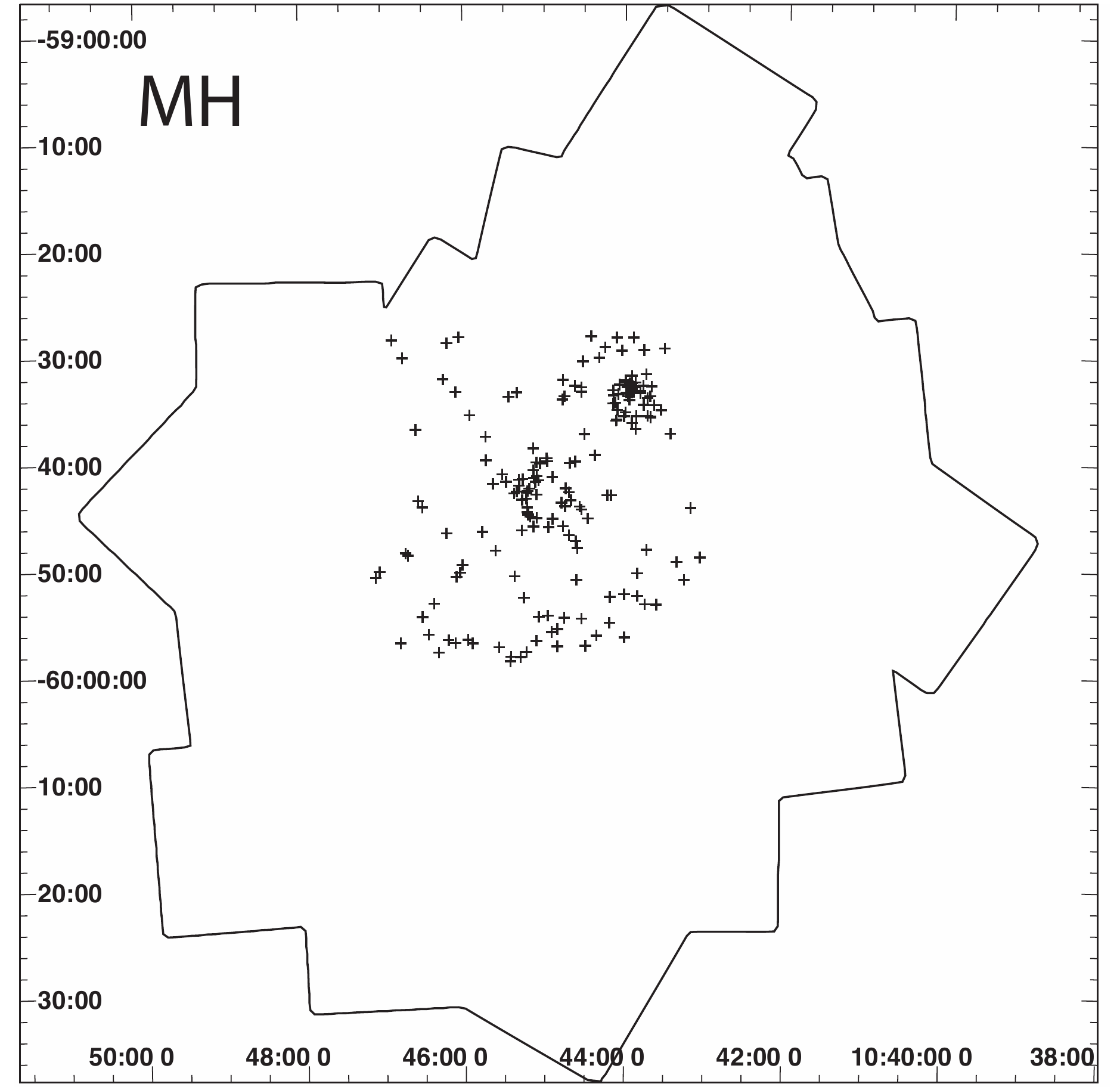}\\
 
\vspace{0.5\baselineskip} 
\includegraphics[width=0.31\textwidth]{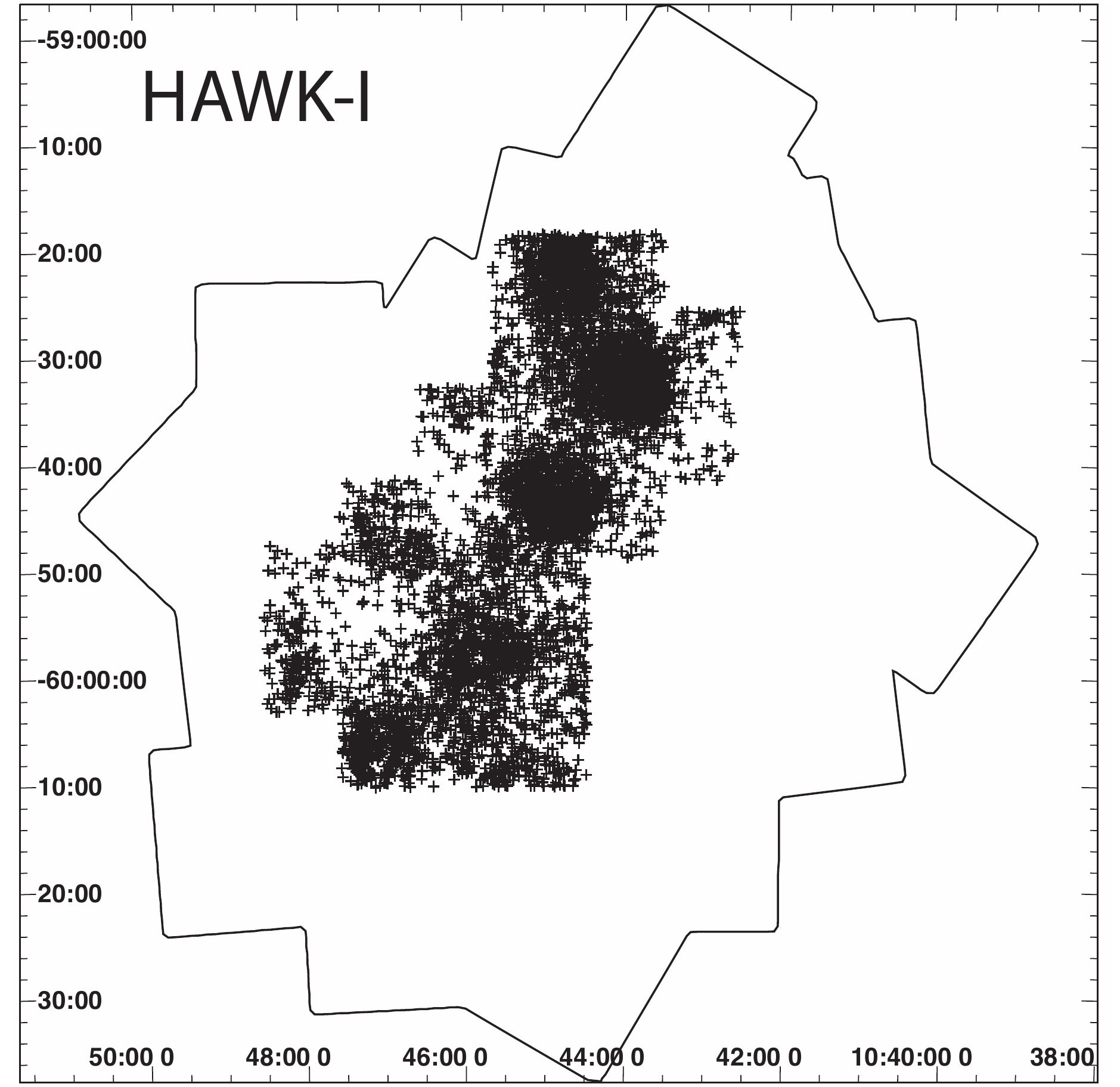}
\includegraphics[width=0.31\textwidth]{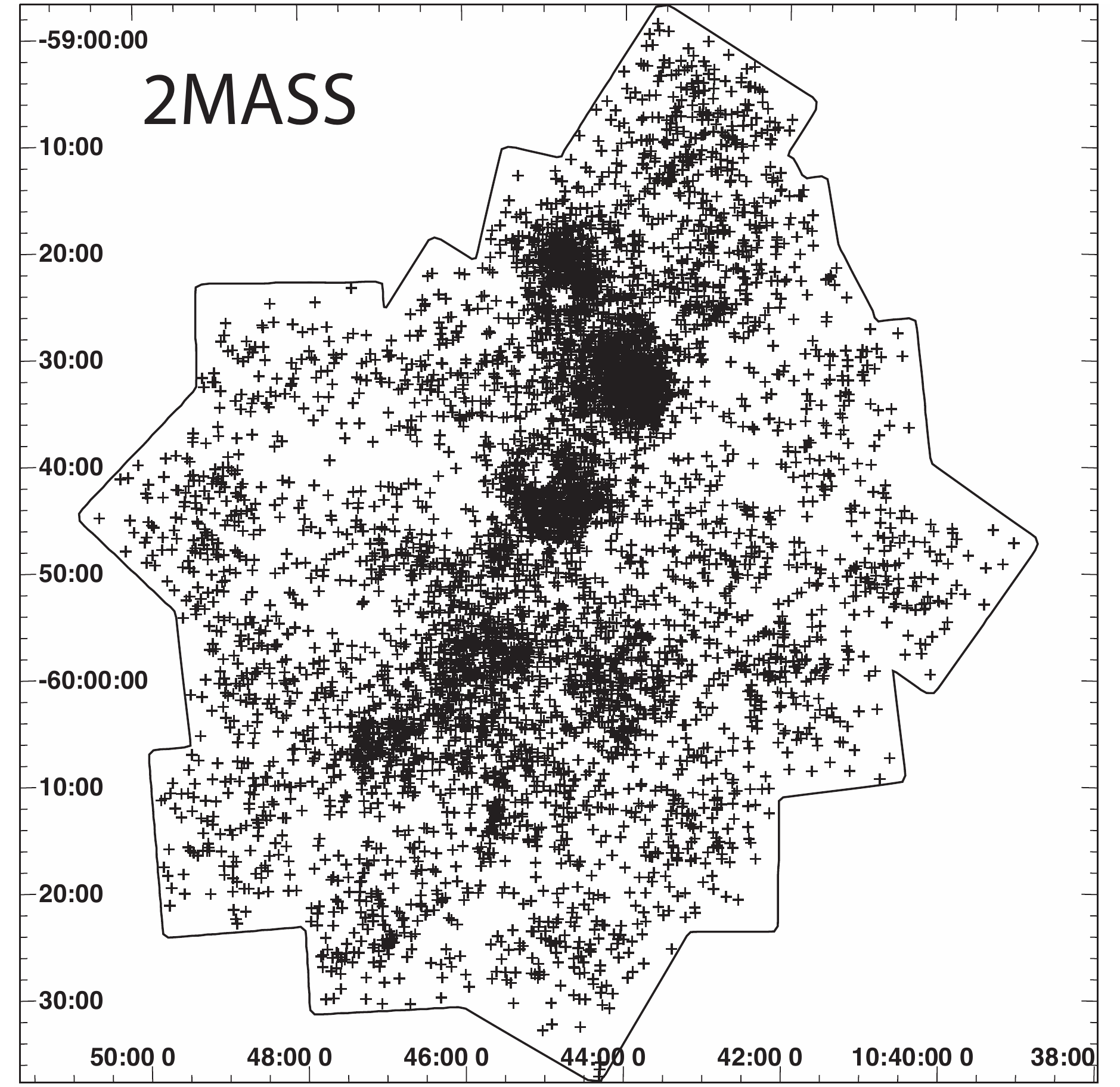}
\includegraphics[width=0.31\textwidth]{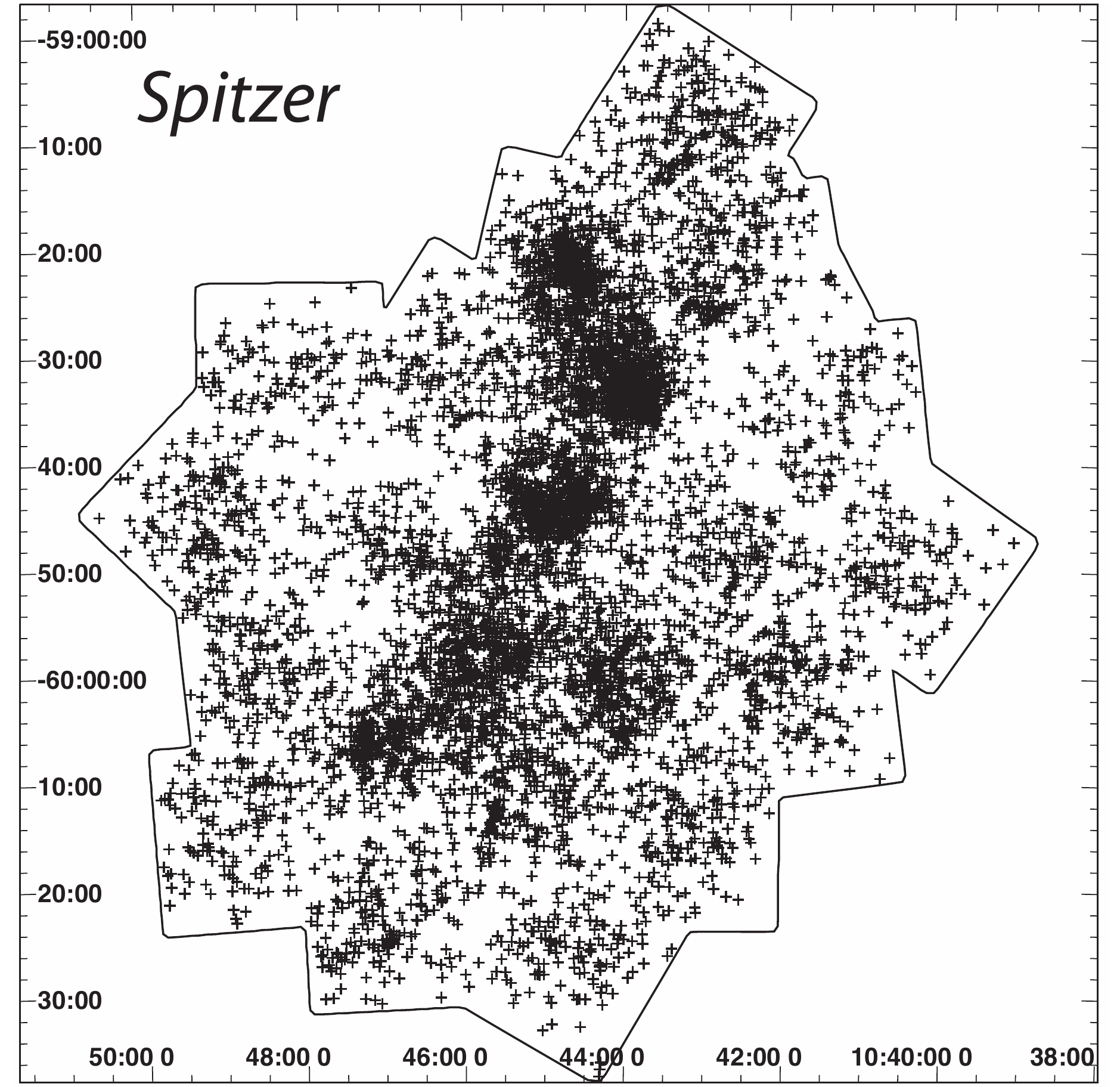}\\
\caption{CCCP matches to selected catalogs in Table~\ref{cat_sum1.tbl}.
The electronic version of the figure can reveal, when zoomed, detail not visible in most printed versions.
Coordinates here and for all subsequent images are celestial J2000.
\label{footprints.fig}
}
\end{figure}

For each counterpart catalog in Table~\ref{cat_sum1.tbl}, $N_{cat}$ gives the approximate number of catalog entries lying in the CCCP field, and $N_{CCCP}$ gives the number of catalog entries that are successfully matched with X-ray sources.   
Preliminary matching runs were used to estimate and remove shifts (but not rotations) between each counterpart catalog and an appropriate astrometric reference, usually the \anchorfoot{http://www.nofs.navy.mil/nomad/}{Naval Observatory Merged Astrometric Dataset} \citep[NOMAD,][]{Zacharias04}; the \ACIS\ data have also been aligned to NOMAD.
These catalog offsets are given in columns (6) and (7).
Median separations among the identified counterparts are reported as $r_{\rm median}$ in Table~\ref{cat_sum1.tbl}.
For most of the catalogs, $>$80\% of the counterparts have separations less than 1\arcsec.  
Figure~\ref{footprints.fig} maps the CCCP sources with identified counterparts for some of the catalogs listed in Table~\ref{cat_sum1.tbl}.  

Selected information about the identified counterparts in all the catalogs in Table~\ref{cat_sum1.tbl} are reported in a single table with 14,368 rows representing the CCCP sources.
The table is available in the electronic version of this paper, and at Vizier \citep{Ochsenbein00}.
Column names and descriptions are listed in Table~\ref{counterpart_properties.tbl}.  
The first four columns are reproduced from Table~\ref{xray_properties.tbl} to identify each CCCP source.
The Identifier, SpType, and SpRef columns provide an historical name for the source, report the spectral type we have adopted in the CCCP, and report a reference for that spectral type.
These quantities are taken from \citet{Gagne11} when available, and from \citet{Skiff09} otherwise.

Columns named ``Offset(*)'' report distances between CCCP and counterpart sources, after astrometric alignment of the catalogs.
The remaining columns appear in groups corresponding to each counterpart catalog.
Within each group, a column named ``Id(*)'' reports a unique identifier in the counterpart catalog.
For the convenience of the reader,\footnote
{Ideally, Table~\ref{counterpart_properties.tbl} would report only the counterpart identifiers, and the reader would retrieve whatever set of previously published counterpart columns are desired at the time Table~\ref{counterpart_properties.tbl} is downloaded.
However, no astronomy catalog server known to the authors can perform this ``join'' service.
}
additional columns give photometry or other information from the counterpart catalog.\footnote{
No screening of the re-published columns has been performed, e.g., removal of photometry estimates that are flagged as low quality by the original authors.
Readers interested in counterpart properties for detailed studies are advised to identify CCCP sources in the original catalogs (using the ``Id'' entries in Table~\ref{counterpart_properties.tbl}) and interpret that counterpart data appropriately.
}

\input{counterpart_column_labels}


\subsection{{\it Chandra} detections of Carina stars with visual spectroscopy and photometry   }

The General Catalog of Stellar Spectral Classifications (``Skiff'' in Table~\ref{cat_sum1.tbl}) is an all-sky collection of stars with MK spectral types based on visual spectroscopy gathered from heterogeneous sources in the literature; 271 Skiff sources are found in the CCCP field.  
\citet{Gagne11} and \citet{Naze11} study 200 O and B stars selected from the Skiff list; for a few of these stars we adopted positions and/or spectral types different from those reported by \citet{Skiff09} \citep{Gagne11}.
Three Wolf-Rayet stars in the field are discussed by \citet{Townsley11a}.
Six stars with spectral types A, F, or K are assumed to lie in the foreground \citep{Broos11}.
A considerable fraction of the known OB stars in the region are not detected in the CCCP \citep{Gagne11} and a considerable number of likely OB members are discovered in the CCCP and do not yet have spectroscopic classifications \citep{Povich11a}.
Thus, the Skiff-CCCP counterparts can not be viewed as well-defined or complete in any criteria.



\begin{figure}
\centering
\includegraphics[width=0.7\textwidth]{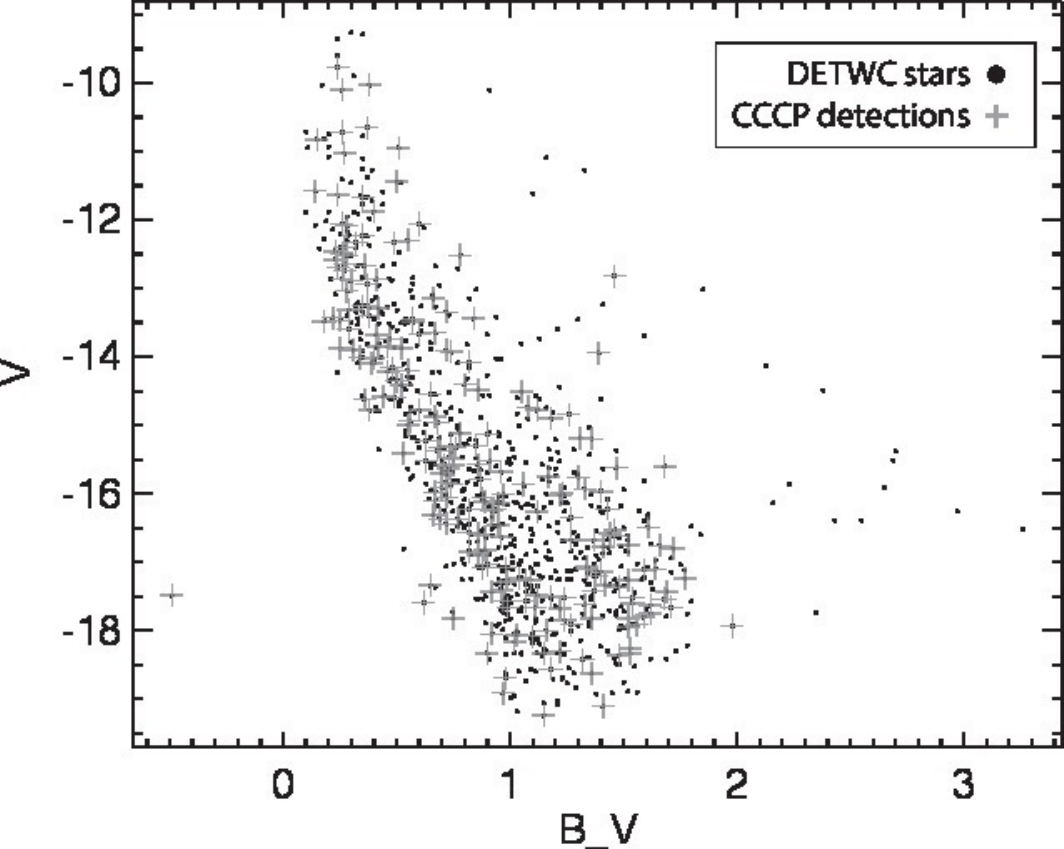} \\
\caption{$BV$ color-magnitude diagram of $V<18$ stars around Tr 14 and Tr 16 reported by \citet{DeGioia01}.  All DETWC sources are shown as dots; CCCP detections are shown as pluses.  \label{DETWC_CMD.fig}}
\end{figure}

\citet{Kharchenko09} (KR) provide an all-sky catalog of visual-band measurements compiled from the literature for more than 2.5 million stars brighter than V $\simeq$ 14 mag; we identify 114 of these in the CCCP catalog.
\citet{Massey93} (MJ) and \citet{DeGioia01} (DETWC) are historically important $UBV$ photometric surveys of the  Tr~16 and Tr~14 regions.   Figure~\ref{DETWC_CMD.fig} shows the color-magnitude diagram for the deeper DETWC survey.   It shows that the CCCP X-ray survey detects a roughly constant fraction of the $18 < V< 12$ DETWC stars at all magnitudes and (moderate) absorptions, from $K$ through $O$ stars.   It is likely that most other CCCP Carina members have similar visual photometric properties.  

Two deep optical $UBVRI$ photometric surveys have been made of a portion of the widely-dispersed Col~228 cluster around 10$^h$43$^m$-60$^\circ$00$\arcmin$\/ with no strong concentration of OB stars.  \citet{Carraro01} (CP) located 1112 stars in a $\sim 20$ arcmin$^2$ field, and \citet{Delgado07} (DAY) located a larger sample in a $\sim 100$ arcmin$^2$ of which 152 are identified as likely pre-main sequence cluster members.     Unfortunately, this location lies off-axis between two ACIS pointings where the X-ray sensitivity is reduced.  Only 27 of the CP stars and 17 of the DAY stars are detected in X-rays.   While small samples, they provide a glimpse at the lower mass function of the stellar population in the southwestern part of the Carina complex.

\subsection{Proper motion measurements of {\it Chandra} stars } \label{proper_motion.sec}

The dense, all-sky PPMXL \citep{Roeser10}  and UCAC3 \citep{Zacharias10} (with its Bright Star Supplement) proper motion surveys have counterparts to 1,006 and 1,471 CCCP sources, respectively.  The Carina proper motion survey of \citet{Cudworth93} (CMD), based on a century of photographic plates, has 141 CCCP counterparts.  

In principle, proper motion measurements can help discriminate rapidly-moving foreground Galactic field stars from Carina members.  
For example, almost all of the PPMXL sources that are spectroscopically confirmed OB stars \citep{Skiff09}---objects very likely to be Carina members---are found to lie in a small circular region of proper motion space\footnote
{                                                                                   
A circular region centered at $(\mu_\alpha,\mu_\delta) = (-6.45, +2.61) $ mas yr$^{-1}$ with a radius of 13 mas yr$^{-1}$ contains 109 of the 110 PPMXL sources that are spectroscopically confirmed OB stars.                                  
}
that contains a minority (445/1006) of all PPMXL sources identified in the CCCP catalog.
However, we were unable to identify a boundary in proper motion space that would select a set of CCCP sources that is clearly consistent with a foreground population, i.e., an unclustered sample with median X-ray energies indicating low interstellar absorption.
This is difficult to interpret, and may reflect observational limitations in automatically measuring proper motions in nebular regions. As we find no simple interpretation to the proper motion characteristics of CCCP sources, we do not consider proper motion to be a useful measurement for classifying CCCP sources \citep{Broos11}.

\subsection{{\it Chandra} counterparts for near-infrared surveys}

Near-infrared ($JHK$) photometry provides vital information to many CCCP studies.
The ubiquitous Two Micron All Sky Survey (2MASS) offers very well calibrated positions and photometry across the entire sky, however its moderate spatial resolution and sensitivity are not adequate to detect CCCP sources that are crowded and/or faint.
For the majority of CCCP sources (those in the major clusters) excellent near-IR data from the High Acuity Wide-field K-band Imager (HAWK-I) instrument\footnote
{
HAWK-I observations were obtained on the ESO 8-meter Very Large Telescope (VLT) at Paranal Observatory, Chile, under ESO programme 60.A-9284(K).
}
\citep{Kissler-Patig08} was available to us; \citet{Preibisch11} discuss these data in depth.
Very high resolution, but very narrow-field, observations of the core of the Tr~14 cluster were provided by \citet{Ascenso07} and by \citet{Sana10}.

\subsection{{\it Chandra} counterparts for wide-field mid-infrared surveys}

Photometric observations of mid-infrared (MIR) point sources are invaluable to studies of star forming regions in a number of ways.
Young stellar objects are readily identified by the MIR emission of their circumstellar disks or accreting envelopes, arising from the reprocessing of stellar radiation.
Strong constraints on the luminosity of OB stars can be obtained from modeling of visual (UBV) and IR spectral energy distributions (SED's), allowing the detection of candidate massive stars and the direct measurement of visual extinction \citep{Povich11a}.
The CCCP had access to two data sets from the {\em Spitzer Space Telescope}: deep observations of two large fields (Spitzer\ Proposal ID 3420, PI N. Smith) covering the South Pillars and the western wall of the southern superbubble lobe \citep{Smith10}, and the shallower Vela-Carina Survey (\Spitzer\ Proposal ID 40791, PI S. Majewski) covering the entire field \citep{Povich11b}.

\section{RELIABILITY OF IDENTIFIED COUNTERPARTS \label{match_outcomes.sec} }

\begin{figure}[htb]
\centering
\includegraphics[width=0.5\textwidth]{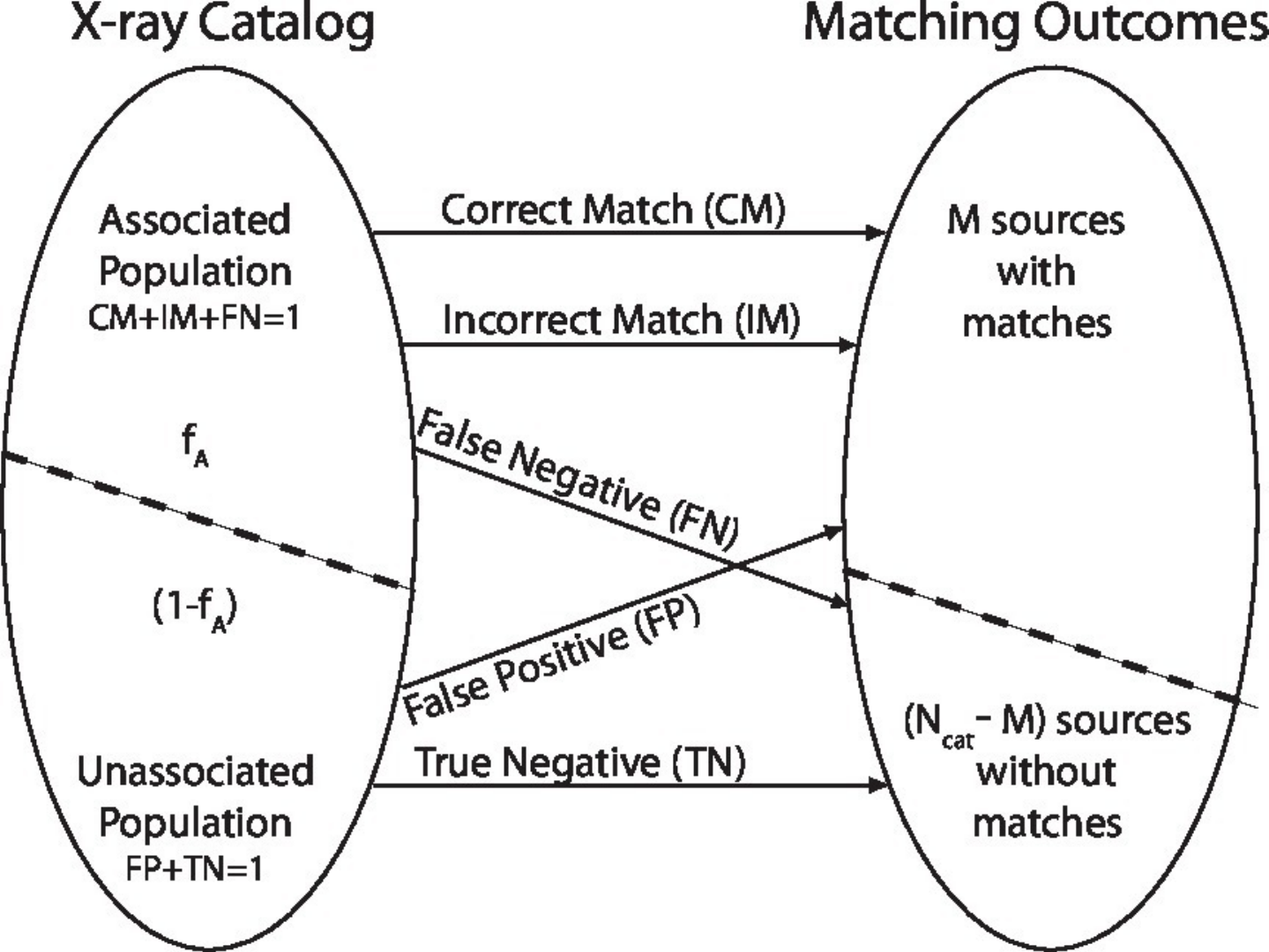} 
\caption{Diagram of all possible ways in which matching an X-ray catalog, comprised of two source populations (left oval), with a counterpart catalog can produce associations (upper portion of right oval) and non-associations (lower portion of right oval).
\S\ref{match_outcomes.sec} defines the outcome pathways (arrows) and their probabilities (CM, IM, FN, FP, TN), and the catalog populations ({\em associated} and {\em unassociated}).
\label{match_xy_simulation.fig}}
\end{figure}

Any counterpart matching enterprise is forced to balance two type of failures---missed matches and spurious matches---analogous to the balance any source detection enterprise must strike between incompleteness and spurious detection. 
\citet[][Appendix]{Broos07} describe a Monte Carlo method for estimating the performance expected from our matching procedure.
The X-ray catalog is modeled as a mixture of two populations: an ``associated population'' for which true counterparts exist in the counterpart catalog, and an ``unassociated population'' for which no counterparts exist in the counterpart catalog.\footnote
{
Some counterparts will be missing from the counterpart catalog, e.g., because they are too bright, too dim, too crowded with other sources, or lost in diffuse emission.
}  
This mixture is represented by the left-hand oval in Figure~\ref{match_xy_simulation.fig}.

In our Monte Carlo method, the process of matching the associated population to the counterpart catalog is simulated many times in order to estimate the frequency of three possible outcomes---correct matches (CM), incorrect matches (IM), and false negatives (FN).  
An IM occurs when the true counterpart is identified as a plausible match, but is rejected in favor of an ``interloper''---another source in the counterpart catalog that (by chance) produces a more plausible match that is mistakenly accepted.  
A FN is a catalogued counterpart that is in fact associated with the X-ray source, but is not identified because the observed X-ray and counterpart positions of the source are too distant to meet the match criterion.
Separate simulations for the unassociated population estimate the frequency of two possible outcomes---true negatives (TN) and false positives (FP).
A FP is a chance superposition of an unrelated source in the counterpart catalog and an X-ray source that has no true counterpart.

The fraction of sources in the actual X-ray catalog expected to belong to the associated population, $f_A$, is estimated by equating the number of actual sources for which no match was identified with the sum of our predictions for the two pathways that can produce that outcome (the FN fraction multiplied by the size of the associated population plus the TN fraction multiplied by the size of the unassociated population):
\begin{equation}
(N_{cat} - M) = N_{cat} [ f_A \times {\rm FN} + (1-f_A) \times {\rm TN} ],
\end{equation}
where $N_{cat}$ is the number of entries in the counterpart catalog and $M$ is the number of X-ray sources that produced matches (see Figure~\ref{match_xy_simulation.fig}).

\begin{figure}[htb]
\centering
\includegraphics[width=1.0\textwidth]{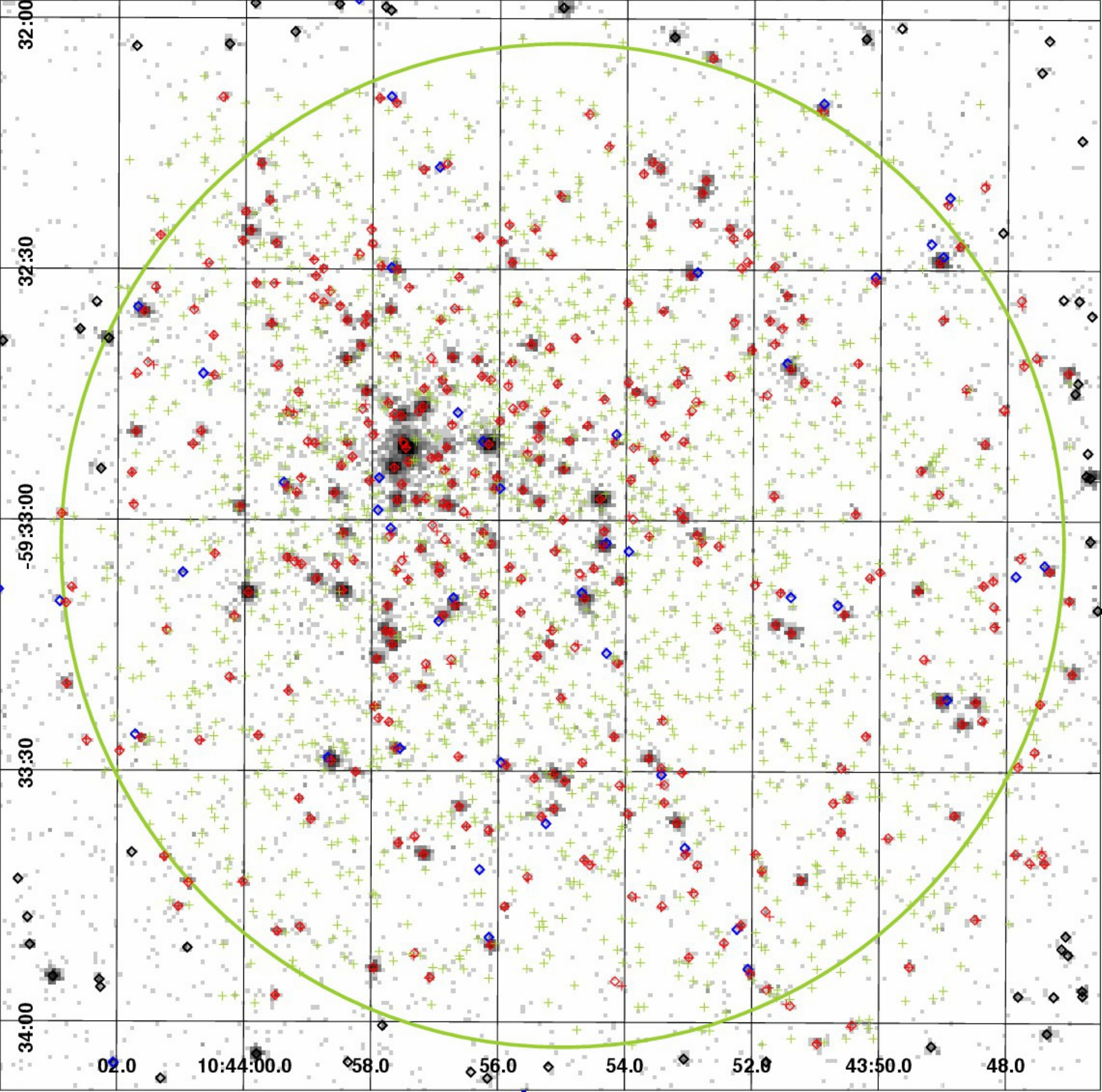}
\caption{Matches identified between the CCCP (red diamonds) and a high-resolution observation of the Tr~14 core \citep{Sana10} (red pluses), with unmatched CCCP (blue diamonds) and Sana (green pluses) sources.
CCCP sources likely to be outside the Sana field of view (green circle) are shown as black diamonds.
The underlying image is the CCCP data, scaled so that pixels with a single X-ray event are light gray.
The electronic version of the figure can reveal, when zoomed, detail not visible in most printed versions.
\todo{Note to referee and journal staff: we would like to examine proofs of this figure before deciding whether a second panel, showing a close-up of the most crowded region, should be added.}
\notetoeditor{We would like to examine proofs of this figure before deciding whether a second panel, showing a close-up of the most crowded region, should be added.}
\label{ACIS-Sana10.fig}}
\end{figure}

We have applied this method to estimate the expected outcomes for the \ACIS/Sana, \ACIS/2MASS, and \ACIS/HAWK-I matching procedures for two complementary samples of the \ACIS\ catalog: a set of very significant ``primary'' sources, defined as having a low probability of being a spurious source (ProbNoSrc\_min in Table~\ref{xray_properties.tbl} less than 0.003) and a set of less significant ``tentative'' sources (\revise{0.003 $<$ ProbNoSrc\_min $<$ 0.01}).
The very deep Sana catalog covers a small field on Tr~14; Figure~\ref{ACIS-Sana10.fig} shows the very good agreement between the CCCP catalog and IR observations in this very crowded field.
The deep HAWK-I catalog covers about one third of the CCCP field (Figure~\ref{footprints.fig}).
Although the 2MASS catalog spans the entire CCCP field of view, we estimated matching outcomes over only the HAWK-I field of view to facilitate comparison with the HAWK-I results.

For the primary and tentative samples and for each counterpart catalog, Table~\ref{match_quality.tbl} reports the total number of \ACIS\ sources (column $N_{CCCP}$) in the field of view analyzed and the expected matching outcomes for those sources: correct matches (CM), incorrect matches (IM), and false negatives (FN) for the associated population; true negatives (TN) and false positives (FP) for the unassociated population.

\begin{deluxetable}{lrrrrrr}
\tablecaption{Estimated outcomes of \ACIS/Sana, \ACIS/HAWK-I, and \ACIS/2MASS matching. \label{match_quality.tbl}                              
}
\tablewidth{0pt}
\tabletypesize{\footnotesize}

\tablehead{
\multicolumn{2}{c}{CCCP Sample}             &  \multicolumn{3}{c}{Associated}            & \multicolumn{2}{c}{Unassociated}   \\
\multicolumn{2}{c}{\hrulefill}              &  \multicolumn{3}{c}{\hrulefill}            & \multicolumn{2}{c}{\hrulefill} \\
\colhead{Reliability} & \colhead{$N_{CCCP}$} & \colhead{CM} & \colhead{IM} & \colhead{FN} & \colhead{TN} & \colhead{FP}   \\
\numberthecolumn & \numberthecolumn & \numberthecolumn & \numberthecolumn & \numberthecolumn & \numberthecolumn & \numberthecolumn    
\setcounter{column_number}{1}
}
\startdata
 & \multicolumn{6}{c}{Sana} \\
primary   &  380 & 81\% &  6\% &  1\%   &  8\% &  3\% \\
tentative &   22 & 38\% &  3\% &  0\%   & 36\% & 22\% \\
\hline
 & \multicolumn{6}{c}{HAWK-I} \\
primary   & 6749 & 74\% &  9\% &  1\%   &  9\% &  7\% \\
tentative &  663 & 23\% &  5\% &  0\%   & 29\% & 43\% \\
\hline
 & \multicolumn{6}{c}{2MASS in HAWK-I field} \\
primary   & 6749 & 50\% &  0\% &  1\%   & 48\% &  2\% \\
tentative &  663 & 12\% &  0\% &  0\%   & 83\% &  4\% 
\enddata
\tablecomments{
Matching outcomes (\S\ref{match_outcomes.sec}) consist of correct matches (CM), incorrect matches (IM), and false negatives (FN) for the associated population; true negatives (TN) and false positives (FP) for the unassociated population. 
}
\end{deluxetable}

As expected, the fraction of \ACIS\ sources (column CM) expected to produce matches that identify {\em true counterparts} varies with the sensitivity of the counterpart catalog: 81\% (primary) and 38\% (tentative) for Sana, 74\% and 23\% for HAWK-I, 50\% and 12\% for 2MASS.  
Among the primary \ACIS\ sources, the fraction of asserted matches expected to be spurious is low: 0.10 for Sana ($[6+3]/[81+6+3]$), 0.18 for HAWK-I ($[9+7]/[74+9+7]$), and 0.04 for 2MASS ($[0+2]/[50+0+2]$).

Among the tentative \ACIS\ sources, spurious matches are expected to be more common.  
Spurious match fractions are 0.40 for Sana ($[3+22]/[38+3+22]$), 0.68 for HAWK-I ($[5+43]/[23+5+43]$), and 0.25 for 2MASS ($[0+4]/[12+0+4]$).
At least three effects can explain this result.  
First, the position errors assigned to the tentative \ACIS\ sources (median 0.44\arcsec) are larger than for the primary sources (median 0.33\arcsec), directly increasing the ``footprints'' around \ACIS\ sources in which a spurious counterpart can be identified.
Second, by definition the tentative sample is expected to harbor a higher fraction of spurious \ACIS\ detections, for which there is no true counterpart to be found.  
Third, since tentative sources tend to be less X-ray luminous than primary sources, if there is any correlation between X-ray and visual/IR luminosity then tentative X-ray sources are likely to have weak visual/IR emission that is not detectable in the counterpart catalog.

\begin{figure}[htb]
\centering
\includegraphics[width=0.5\textwidth]{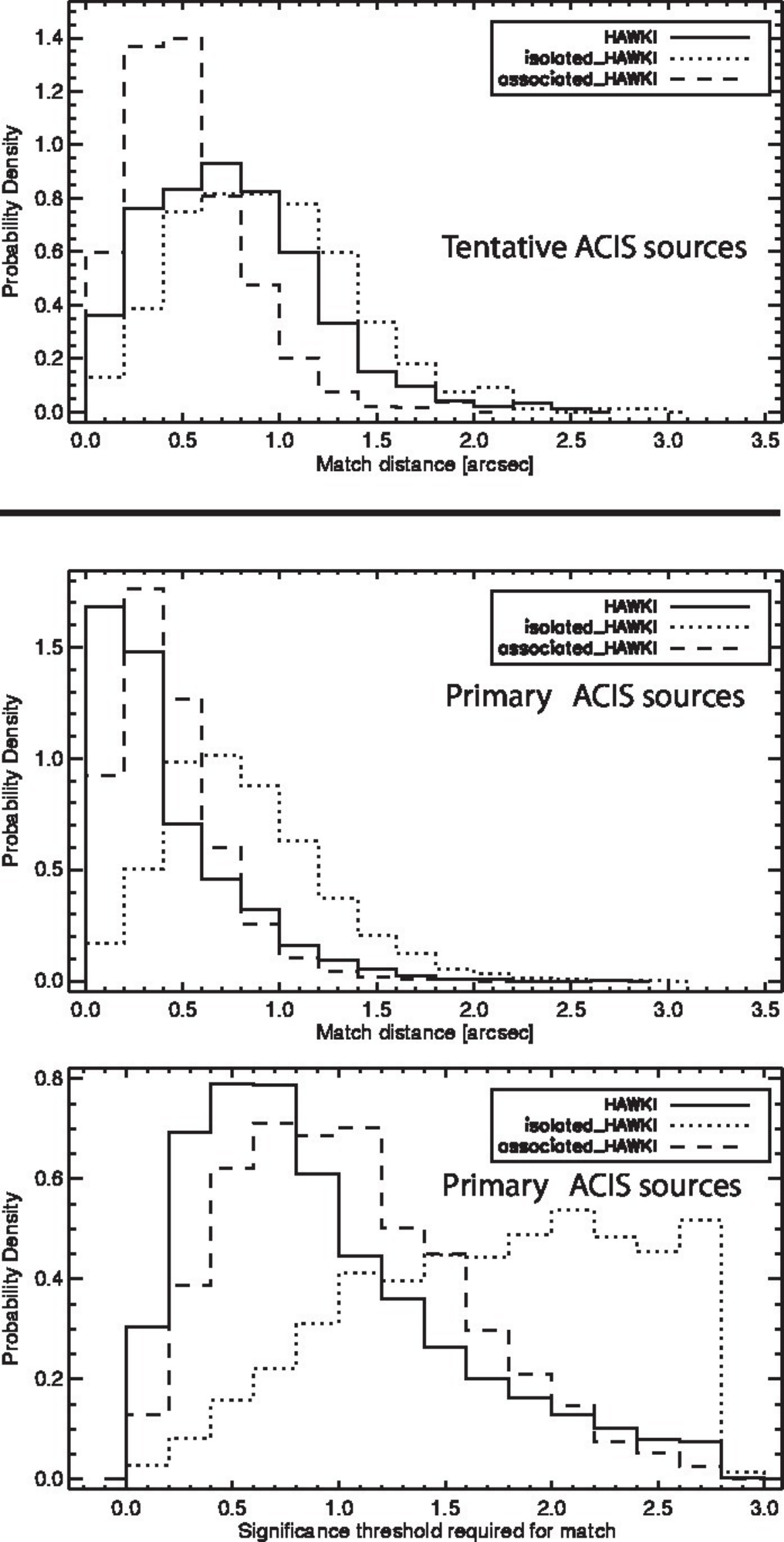}
\caption{Observed separations (solid histograms) between tentative (upper panel) and primary (middle and lower panels) \ACIS\ sources and HAWK-I counterparts, compared with simulated separations for an associated population (dashed) and an unassociated population (dotted). 
Separations are characterized via angular distance (upper and middle panels) and via match significance (lower panel).
\label{match_distance_simulated.fig}}
\end{figure}

The notion that the X-ray catalog can be viewed as a mixture of associated and unassociated populations---on which the matching outcome estimation method above is based---is visually demonstrated in Figure~\ref{match_distance_simulated.fig}, where \ACIS/HAWK-I separations (in arcseconds) of matching sources are shown for the two simulated populations and for the actual \ACIS\ catalog.
The upper panel shows that the tentative \ACIS\ catalog (solid) has match separations that appear to be a mixture of the associated population (dashed) and the unassociated population (dotted).
The middle panel shows the same quantities for the primary \ACIS\ catalog; the fact that the actual \ACIS/HAWK-I  separations (solid) are smaller than those from the associated simulation (dashed) suggests that the adopted \ACIS\ and/or HAWK-I position errors are over-estimated.
Perhaps a more informative measure of source separation is the smallest significance threshold, a parameter to the matching algorithm \citep[see][\S8]{Broos10}, required to assert the match.
The bottom panel presents the distribution of this quantity (expressed as a multiple of $\sigma$ for a Gaussian distribution) for the same catalogs and simulations as the middle panel.
Separations among matches for the associated population (dashed) most commonly fall at ${\sim}1\sigma$, falling off at higher separations.
In contrast, the occurrence of separations in the unassociated population (dotted) increases with separation, reflecting the quadratic growth in area of annular footprints with increasing distance from the \ACIS\ source. 
As in the middle panel, the actual \ACIS/HAWK-I matches (solid) are generally smaller than those from the associated simulation (dashed), suggesting that the adopted \ACIS\ and/or HAWK-I position errors are over-estimated.

\section{QUALITY OF THE X-RAY CATALOG \label{quality.sec}}

\subsection{Sensitivity to Point Sources  \label{sensitivity.sec}}

We find that very precise language is required to avoid miscommunication when discussing the ``sensitivity'' of a source catalog.
Here, we restrict our use of that word to mean the general concept that, for any given observation, the probability of detecting a source depends on how ``bright'' it is;  we do not use the word ``sensitivity'' to refer to any specific threshold.
We use the term ``completeness'' to refer to the probability that members of a precisely defined parent population of point-like objects will be present in a precisely defined sample of detected sources.
In contexts where the parent population under discussion is defined by some observational or astrophysical quantity, we use the term ``completeness limit'' to mean a lower-limit on that quantity for which nearly 100\% of the parent population will be present in a specified sample.
Thus, a ``completeness limit'' is properly stated in terms of a specific measured or inferred source property and in terms of a specific sample of detected sources.

Most studies that involve X-ray-selected sources will have to consider to what extent the analysis is affected by the finite sensitivity of the X-ray observations.
Sensitivity is perhaps best studied via Monte Carlo simulations of synthetic data sets that contain artificial sources that are subjected to the same detection procedure used on the observed data.
Since most studies need to characterize sensitivity in terms of an astrophysical, not an observational, quantity such as intrinsic (corrected for absorption) X-ray luminosity or stellar mass, the first simulation task is astrophysical---modeling the X-ray photons that \Chandra\ should see for various astrophysical objects.
This task is often quite difficult, requiring information that is not well known (e.g., the 3-D distribution of absorbing material and the 3-D distribution of sources) and/or requiring astrophysical models that are not well known (e.g., the relationship between stellar mass and X-ray luminosity).

After astrophysical models have predicted {\em apparent} (diminished by absorption) X-ray flux for synthetic stars, the second simulation task is instrumental---modeling the event data (including background) that would be produced, the data reduction process, and the source detection process.
If the layout of a \Chandra\ mosaic (exposure times and the pattern of observation overlap) is simple and the source detection procedure is simple, then apparent flux sensitivity can be studied without detailed Monte Carlo simulations; see for example \citet{Georgakakis08} and references therein.
However, for the complex data set and the complex detection procedure employed in the CCCP, adequate resources are simply not available to perform the complex simulations that would be required to map detection probabilities across the field. 

Instead, we aspire to present here only an empirical description of detection completeness limits with respect to apparent photometric quantities, based on the astrophysical assumption that those quantities follow a powerlaw distribution that ``rolls off'' at the low-flux end due to decreasing detection probability. 
Our primary goal is to visualize and quantify the large influence that off-axis angle (Theta in Table~\ref{xray_properties.tbl}) has on sensitivity in the CCCP catalog (and indeed in many other \Chandra\ catalogs).
This relationship arises from two instrumental effects.
First, the extraction area, and thus any spatially-uniform background component, of a point source with a reasonable aperture (e.g., 90\% PSF fraction) is ${\sim}$100 times larger far off-axis ($\theta = 10$\arcmin) than on-axis.
Second, the \Chandra\ effective area far off-axis 
is ${\sim}$20\% lower than on-axis due to mirror vignetting.\footnote
{See Figure~4.5 in the \anchorparen{http://asc.harvard.edu/proposer/POG/pog_pdf.html}{\Chandra\ Proposers' Observatory Guide}.}

\begin{figure}[htb]
\centering
\includegraphics[width=0.5\textwidth]{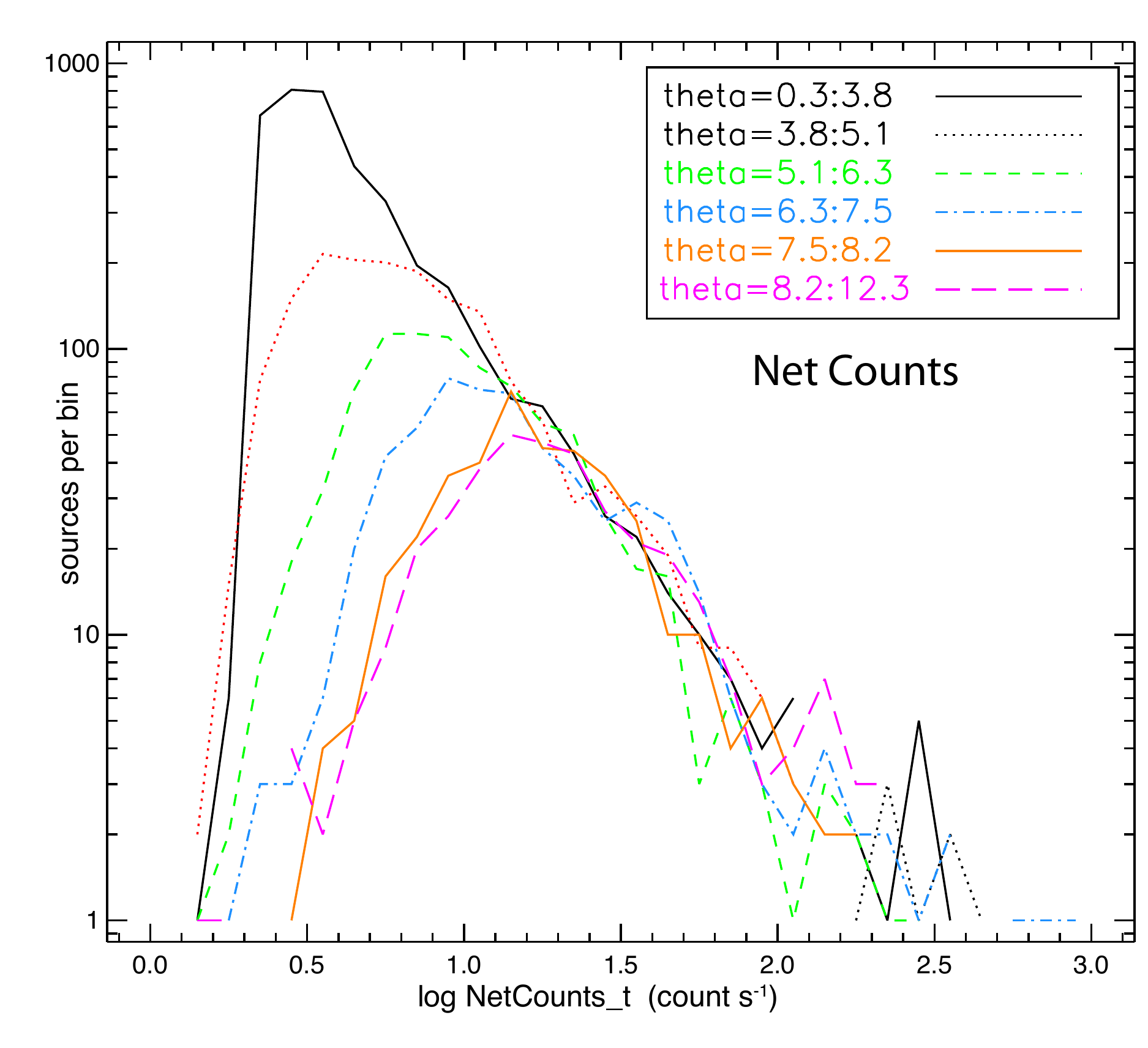} \\
\includegraphics[width=0.5\textwidth]{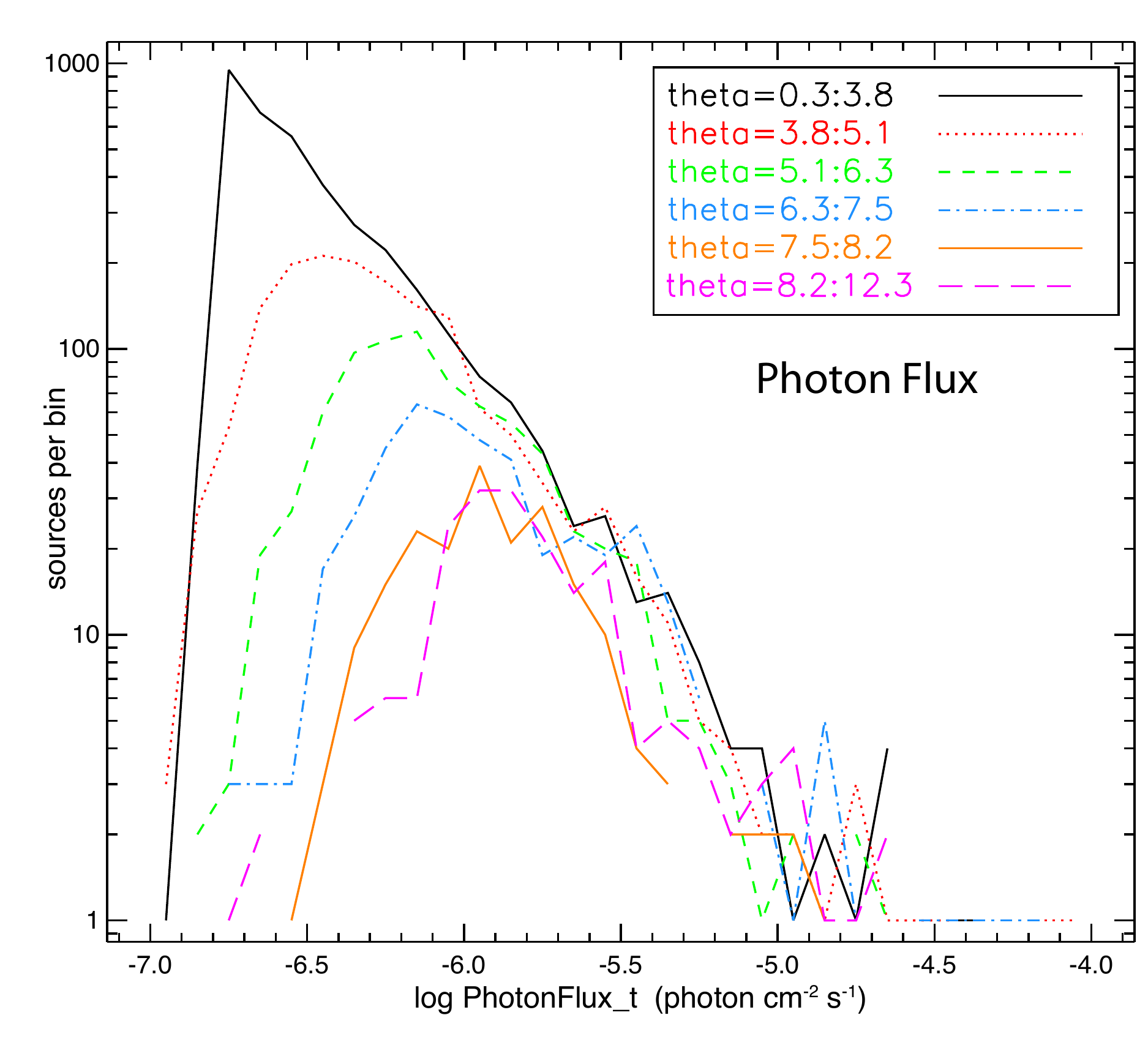}
\caption{Histogram of net extracted counts (upper panel) and apparent photon flux (lower panel) for a sample of CCCP sources defined in \S\ref{sensitivity.sec}, stratified by off-axis angle.
\label{theta_slices.fig}}
\end{figure}

\begin{figure}[htb]
\centering
\includegraphics[width=0.5\textwidth]{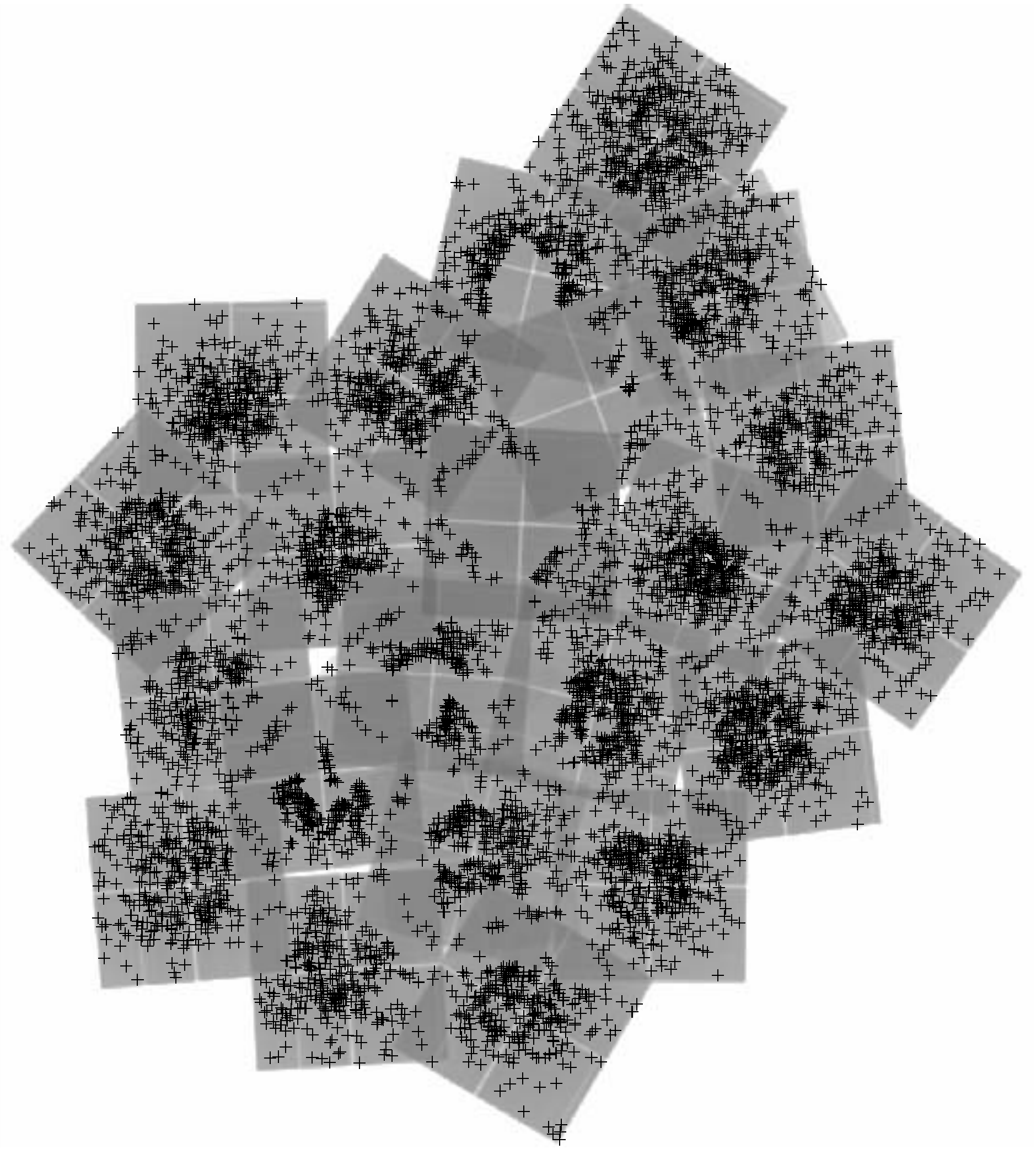}
\caption{CCCP sample used in Figure~\ref{theta_slices.fig} and to estimate completeness limits in Table~\ref{theta_slices.tbl}; these are sources with well-defined off-axis angle that are not obviously associated with clusters.
\label{NetCounts_sample.fig}}
\end{figure}

The upper panel of Figure~\ref{theta_slices.fig} shows histograms of the observational quantity NetCounts\_t from Table~\ref{xray_properties.tbl} (net extracted counts in the total band, 0.5--8~keV) over six disjoint ranges of off-axis angle. 
The source sample represented in Fig.~\ref{theta_slices.fig} is mapped in Figure~\ref{NetCounts_sample.fig}; we excluded 6877 sources that may exhibit intrinsic spatial variation in the luminosity function (those lying in clusters identified by \citet{Feigelson11}) or that have a poorly-defined off-axis angle (more than 20\% variation in Theta among the covering observations).

We choose to interpret the empirical distribution of NetCounts\_t in each off-axis angle slice as arising from a powerlaw-distributed parent population of astrophysical sources (young stars in Carina, field stars, extragalactic objects) that has been incompletely detected at low fluxes.
We performed a maximum likelihood fit of the unbinned NetCounts\_t data in each slice to a powerlaw model with a low-side cutoff\footnote
{
\citet{Maschberger09} discuss the well-known unbiased maximum likelihood estimator for the exponent of an infinite powerlaw distribution and 
they propose a similar unbiased estimator for a truncated powerlaw.
We have implemented these estimators using IDL in a tool called {\em ml\_powerlaw}, which is part of our \anchor{http://www.astro.psu.edu/xray/acis/acis_analysis.html}{\TARA} software package (\url{http://www.astro.psu.edu/xray/acis/acis_analysis.html}).
}, and we choose to interpret that cutoff value as the onset of incompleteness.
The best-fit powerlaw exponent and the inferred completeness limit are reported as $\alpha$ and $N_{\rm lim}$ in the middle section of  Table~\ref{theta_slices.tbl}.

\begin{deluxetable}{lcrccrccc}
\tablecaption{Completeness Limits for the CCCP Catalog \label{theta_slices.tbl}
}
\tablewidth{0pt}
\tabletypesize{\footnotesize}

\tablehead{
\multicolumn{2}{c}{Off-axis Angle} & \multicolumn{3}{c}{NetCounts\_t (0.5--8 keV)} & \multicolumn{4}{c}{$F_{\rm t,photon}$ (0.5--8 keV)} \\ 
\multicolumn{2}{c}{\hrulefill}     & \multicolumn{3}{c}{\hrulefill}   & \multicolumn{4}{c}{\hrulefill}    \\
\colhead{Range} & \colhead{Median} & 
\colhead{$N_{sample}$} & \colhead{$\alpha$} & \colhead{$\log N_{\rm lim}$}    &  
\colhead{$N_{sample}$} & \colhead{$\alpha$} & \colhead{$\log F_{\rm t,photon,lim}$} & \colhead{$\log L_{\rm t,c,lim}$}  \\
\colhead{(\arcmin)} & \colhead{(\arcmin)} & \colhead{(source)} & &\colhead{(count)} & \colhead{(source)} & & \colhead{(photon~cm$^{-2}$~s$^{-1}$)}  & \colhead{(erg~s$^{-1}$)} \\
\numberthecolumn & \numberthecolumn & \numberthecolumn & \numberthecolumn & \numberthecolumn & \numberthecolumn & \numberthecolumn & \numberthecolumn & \numberthecolumn  
\setcounter{column_number}{1}
}
\startdata
0.0:3.8   & 2.8  & 3766~~ & 2.41 & 0.8  & 3651~~ & 2.29 & -6.7 & 29.9  \\ 
3.8:5.1   & 4.4  & 1613~~ & 2.38 & 0.9  & 1523~~ & 2.34 & -6.2 & 30.4  \\ 
5.1:6.3   & 5.7  &  814~~ & 2.58 & 1.2  &  749~~ & 2.34 & -6.2 & 30.4  \\ 
6.3:7.5   & 6.9  &  548~~ & 2.24 & 1.2  &  426~~ & 2.11 & -6.1 & 30.5  \\ 
7.5:8.2   & 7.9  &  387~~ & 2.40 & 1.2  &  198~~ & 2.62 & -5.9 & 30.7  \\ 
8.2:12.3  & 8.7  &  356~~ & 2.28 & 1.3  &  189~~ & 2.31 & -5.9 & 30.7  \\ 
\enddata
\tablecomments{~\\
Col.\ (1): Off-axis angle range defining the sample
\\Col.\ (2): Median off-axis angle within the sample
\\Cols.\ (3) and (6): Sample size
\\Cols.\ (4) and (7): Estimated exponent of truncated powerlaw model
\\Cols.\ (5), (8), and (9): Estimated truncation threshold (completeness limit)
}
\end{deluxetable} 
                                                                                             
The lower panel of Figure~\ref{theta_slices.fig} and right section of Table~\ref{theta_slices.tbl} present the same analysis carried out on a simple {\em calibrated} photometric quantity, an estimate of apparent (not corrected for absorption) photon flux in the total energy band (0.5--8~keV), $F_{\rm t,photon}$ defined by Equation~\ref{photon_flux.eqn}.  
The source sample depicted is the NetCounts\_t sample, less 748 sources that have off-nominal exposure time.
The best-fit powerlaw exponent and the inferred photon flux completeness limit are reported as $\alpha$ and $F_{\rm t,photon,lim}$.

Those photon flux limits have been converted to intrinsic (corrected for absorption) luminosity limits in the total energy band (0.5--8~keV), shown as column $L_{\rm t,c,lim}$, by making the following astrophysical assumptions.
The emitting gas is modeled as the two-temperature thermal plasma that \citet{Getman10} adopt as typical for low-mass pre-main sequence stars with moderate intrinsic luminosity, specifically the model shown in the $L_{h,c} = 10^{30.0}$ entry of their Table~1.
The absorbing column is assumed to be $N_H = 0.64 \times 10^{22}$ cm$^{-2}$ corresponding to $A_V = 4$ mag, which \citet{Preibisch11} find to be typical for Carina stars.
The distance to Carina is assumed to be 2300 pc.

%

We wish to emphasize that the completeness limits in Table~\ref{theta_slices.tbl} are a characterization of source samples defined only by the CCCP detection process and by off-axis angle.
Once a scientific study has imposed additional selection criteria---e.g., source classification, the availability of counterpart information, or the availability of individual X-ray luminosity estimates---the completeness limit for the resulting sample can be higher.

The strong variation in detection completeness with off-axis angle (${\sim}0.8$ dex for photon flux) inferred from Table~\ref{theta_slices.tbl} is responsible for the obvious ``egg-crate effect''---a greater density of sources on-axis than off-axis---seen when the full catalog is plotted on the sky, as in \tbr{Figure~4} in \citet{Townsley11a}.
A sensible method for suppressing this effect from any particular region on the sky (e.g., for a particular star cluster) is to trim the catalog at the photon flux limit corresponding to the largest off-axis angle included in that region.
Such an approach is recommended for spatial analyses involving counting sources e.g., \citet[][]{Feigelson11}.  
In that study, the ``complete'' sample of 3,220 Carina members is defined by $\log F_{\rm t,photon,lim} > -5.9$ photon cm$^{-2}$ s$^{-1}$ based on the $F_{\rm t,photon,lim}$ completeness limits shown in Table~\ref{theta_slices.tbl}. 

However, we must emphasize that a catalog so trimmed is not expected to be complete in {\em other} astrophysical quantities, such as apparent X-ray luminosity (which depends on the temperature of the source) or intrinsic X-ray luminosity (which additionally depends on absorption to the source).
In fact, applying a cut in NetCounts\_t or PhotonFlux\_t is {\em harmful} in such analyses because some of the discarded sources may in fact lie above the completeness limit for the quantity of interest.

Spurious sources have the potential to mask the incompleteness of legitimate detections at low fluxes, thereby biasing estimates of  completeness limits.
For example, in analyses like those in \S\ref{sensitivity.sec}, spurious sources could ``fill in'' the deficit of detected legitimate sources at low fluxes in a way that obscures the departure of the detected distribution from a powerlaw model.
We can roughly assess the degree to which the inferred completeness limits in Table~\ref{theta_slices.tbl} may suffer from this issue by estimating an upper limit on the fraction of sources brighter than those limits that have a reasonable risk of being spurious.
For example, within the innermost off-axis angle slice ($\theta = 0$--3.8\arcmin) in Table~\ref{theta_slices.tbl}, ${\sim}10$\% of the sources above the stated PhotonFlux\_t completeness limit are ``tentative'' detections (defined by \revise{0.003 $<$ ProbNoSrc\_min $<$ 0.01} in Table~\ref{xray_properties.tbl}).
Thus, that completeness limit may be somewhat optimistic.
All other complete samples in that table contain $<$3\% tentative sources.

\subsection{Spurious Sources \label{spurious.sec}}

Our detection process is intentionally aggressive, and thus some weak spurious sources should be expected.
The ultimate strategy for quantifying spurious detections is perhaps Monte Carlo simulation of the detection process executed on synthetic data sets that match the actual observations as closely as possible.
Such simulations require an estimate of the spatially-variable background in the observations, which is problematic because by definition one must decide which photons come from point sources in order to identify which photons are ``background''.
Such simulations must also include artificial point sources because presumably one of the mechanisms for producing a spurious candidate in our detection procedure (\S\ref{extraction.sec}) is imperfect image reconstruction in the wings of bright point sources.
\revise{Such simulations are infeasible due to the complexity of the Carina survey and the complexity of our detection procedures.
Even if they could be performed, the result of such simulations---a characterization of the spurious source population in the full CCCP catalog---would be of little practical use because science analyses invariably involve additional sample selections of various kinds (e.g., matching to counterpart catalogs, source classification, selection within a region of interest on the sky).
Propagating simulated spurious sources through sample selection and into science analysis would be a very difficult task.}

Perhaps the best method for gauging whether spurious sources may impact a scientific conclusion is to repeat the relevant science analysis on a smaller source sample obtained by applying a stricter criterion for source existence, e.g., by applying a smaller threshold on the ProbNoSrc\_min (Table~\ref{xray_properties.tbl}) statistic that was used to define the CCCP catalog (\S\ref{extraction.sec}). 
Often, other criteria that define the source sample under study (e.g., involving counterparts, source classification, completeness cuts) have already removed most sources whose existence is uncertain.  For example, in \S\ref{sensitivity.sec} we found that ``tentative'' detections are virtually absent from samples that are constructed to be complete in apparent photon flux. 

\begin{figure}
\centering
\includegraphics[width=0.7\textwidth]{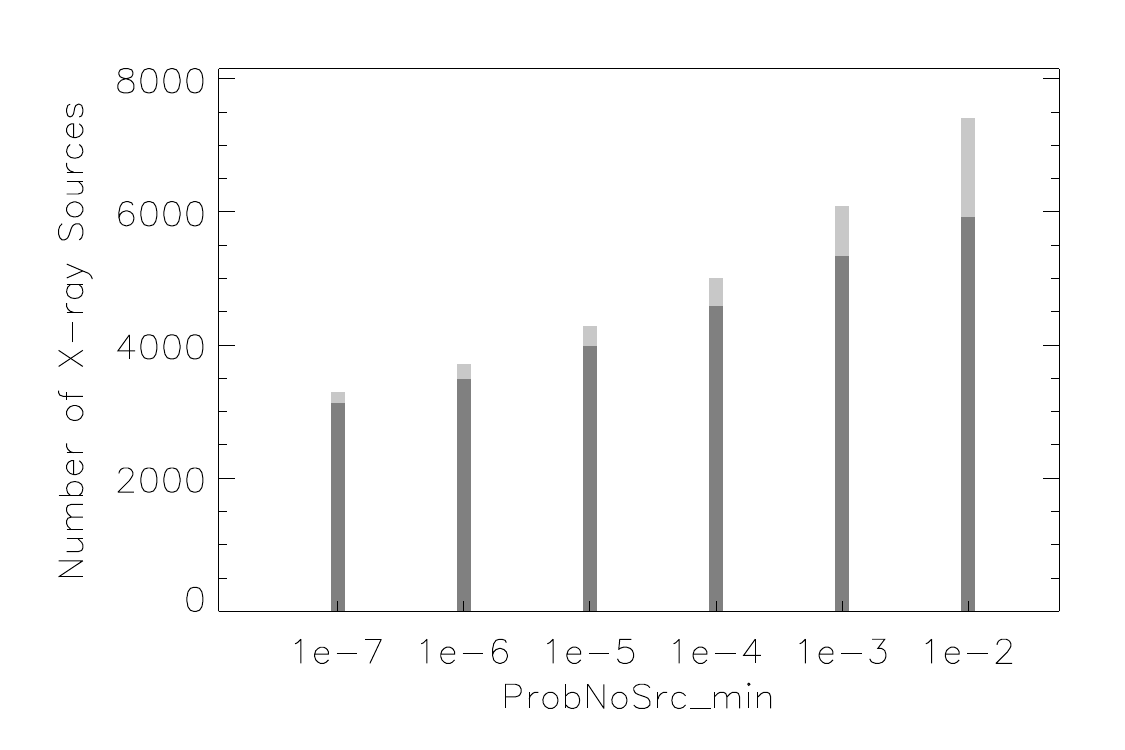} \\
\caption{Estimated size of the associated (dark gray) and unassociated (light gray) components of hypothetical CCCP catalogs (within the HAWK-I field of view) defined by various thresholds on source significance (the ProbNoSrc\_min statistic in \S\ref{extraction.sec}). \label{associated_vs_Pb.fig}}
\end{figure}

\revise{Although we do not have an accurate estimate for the number of spurious sources in the CCCP catalog, a reasonable {\em upper limit} 
for that quantity is the number of X-ray sources for which no {\em true counterpart} exists in the deep HAWK-I infrared catalog, defined as the ``unassociated X-ray population'' in \S\ref{match_outcomes.sec}.
Recall that the size of the unassociated population and the size of its complement, the ``associated population'', are conceptually distinct from the number of X-ray sources without or with identified {\em matches}; the unassociated/associated fractions are estimated via detailed Monte Carlo simulations.

The size of the unassociated population is an upper limit, rather than a fair estimate, on the number of spurious X-ray detections because not every legitimate X-ray source is expected to be detected by HAWK-I.
For example, the ${\sim}700$ extragalactic X-ray sources in the HAWK-I field of view expected to be detected by the CCCP should be very faint in the near infrared \citep{Getman11}; many will be below the completeness limits of HAWKI-I \citep[J${\sim}$21, H${\sim}$20, K${\sim}$19,][]{Preibisch11}.


We have estimated the sizes of the unassociated and associated populations, with respect to HAWK-I, for the full CCCP catalog and for several smaller hypothetical X-ray catalogs (within the HAWK-I field of view) obtained by adopting more conservative requirements on source significance (smaller values of the ProbNoSrc\_min statistic described in \S\ref{extraction.sec}).
Figure~\ref{associated_vs_Pb.fig} shows that the size of the associated population (dark gray bars), which is assumed here to be a lower limit on the size of the legitimate X-ray detections, grows as we accept less and less certain X-ray detections.
At the ProbNoSrc\_min threshold we chose for the CCCP catalog (0.01), the associated population appears to still be growing.
The cost of pushing deeper into the data is, of course, a rise in the number of spurious sources, assumed here to be bounded by the unassociated population (light gray bars).
}

\revise{There is no {\em correct} decision regarding the tradeoff between sensitivity and reliability.
We generally take an aggressive approach to source detection for a variety of reasons.
From a near term perspective, for studies of diffuse emission we feel that the added contamination arising from failing to identify and mask actual point sources is more damaging than the observing area lost to masking spurious point source detections.
From a long term perspective, we are painfully aware that multiple decades may pass before this target is observed again by an X-ray observatory with an angular resolution superior to \Chandra.  
We believe that recording tentative X-ray detections in the literature {\em may} prove valuable to future investigators, who will perhaps have access to observations in other bands that are far superior to what is currently available. 
}

%

\section{SUMMARY}

A careful analysis of the 38 \ACIS-I observations (totaling 1.2 Ms) that comprise the \Chandra\ Carina Complex Project (CCCP) identifies and extracts 14,369 X-ray point sources; their positions and basic X-ray properties are presented in an electronic table.
The catalog appears to be complete across most of the field to a limit of ${\sim}$20 net X-ray counts 
\revise{(corresponding to an absorption-corrected total-band luminosity of ${\sim}10^{30.7}$ erg~s$^{-1}$ for a typical low-mass pre-main sequence star).}
Near the centers of the \ACIS\ pointings, where the \Chandra\ PSF is optimized, the completeness limit improves to ${\sim}$6 net counts 
\revise{(corresponding to an absorption-corrected total-band luminosity of ${\sim}10^{29.9}$ erg~s$^{-1}$ for a typical low-mass pre-main sequence star).}
A large fraction of the detected sources lies beyond these completeness limits, e.g., more than half produced less than 10 net X-ray counts.

Forty sources have high-quality X-ray spectra (more than 500 net counts); most are known \citep{Gagne11,Naze11,Parkin11,Townsley11a} or candidate \citep{Evans11,Povich11a} massive stars.
Five suffer from mild photon pile-up in the detector; a new method for pile-up correction is described in Appendix~\ref{pile-up_recon.sec}.

Counterparts to the majority (69\%) of the X-ray sources are identified in a variety of visual and infrared catalogs, including newly-available wide-field and deep surveys in the near-IR (HAWK-I on the VLT) and mid-IR (\Spitzer) bands. 
Most ($>$80\%) of the X-ray/counterpart source separations are $<$1\arcsec; the reliability of the counterpart identifications is discussed.

The X-ray catalog and counterpart identifications presented here provide the basis for later studies of the young stellar population in the Carina complex.  
Removal of the events associated with these sources allows study of the diffuse emission in this starburst region \citep{Townsley11b}.

\appendix

\section{CHANDRA CCD PILE-UP RECONSTRUCTION \label{pile-up_recon.sec}}

As outlined in \S\ref{catalog.sec}, a few CCCP sources are sufficiently strong X-ray emitters to suffer from \anchorfoot{http://cxc.harvard.edu/ciao/why/pileup_intro.html}{{\em photon pile-up}}---the arrival of multiple X-ray photons with a separation in time and space that is too small to allow each to be detected as a separate X-ray event.
This phenomenon is ubiquitous among CCD-type detectors in X-ray astronomy and prevents standard analysis of source flux, variability, and spectrum \citep{Allen98,Bautz00}.

Various treatments of pile-up for the \Chandra\ \ACIS\ instrument have been proposed.  
\citet{Broos98} and surely other authors point out that pile-up effects in point source spectra can be reduced by simply discarding events in the core of the observatory point spread function, where the incident photon flux is highest, and then applying appropriate energy-dependent calibration corrections to account for the discarded collecting area.
\AEacro\ supports this approach;\footnote{\url{http://www.astro.psu.edu/xray/docs/TARA/ae_users_guide/pileup.txt}} for example, \citet{Getman05} perform annular extractions for dozens of piled sources in the Orion Nebula Cluster.  

The most commonly used pile-up analysis technique involves including pile-up effects in the observatory model employed by ``forward fitting'' spectral analysis packages such as \Sherpa\ and \XSPEC.
These tools hypothesize an astrophysical model of the X-ray source, predict the instrumental spectrum that should be observed by employing observatory calibration information, compare the predicted and observed spectra, and adjust parameters of the astrophysical model to improve the prediction.
\citet{Chartas00} describe the derivation of astrophysical models for piled sources, using \XSPEC\ to implement an astrophysical model and using a CCD simulator \citep{Townsley02a} to model the response of \ACIS, including pile-up effects.
\citet{Davis01} modeled pile-up with a parametrized integral equation, based on work by \citet{Ballet99}.
This model, which we will hereafter refer to as the ``Davis model'', has been implemented in all the spectral analysis packages in common use by \ACIS\ observers, and has been the sole technique applied to most piled \ACIS\ sources for the past decade.
Bayesian statistical methods have also been studied \citep{Yu00a,Yu00b,Kang03,Yu04}, but are not in common usage among observers.

\subsection{A New Forward-Fitting Algorithm}

\begin{figure}[htb]
\centering
\includegraphics[width=0.5\textwidth]{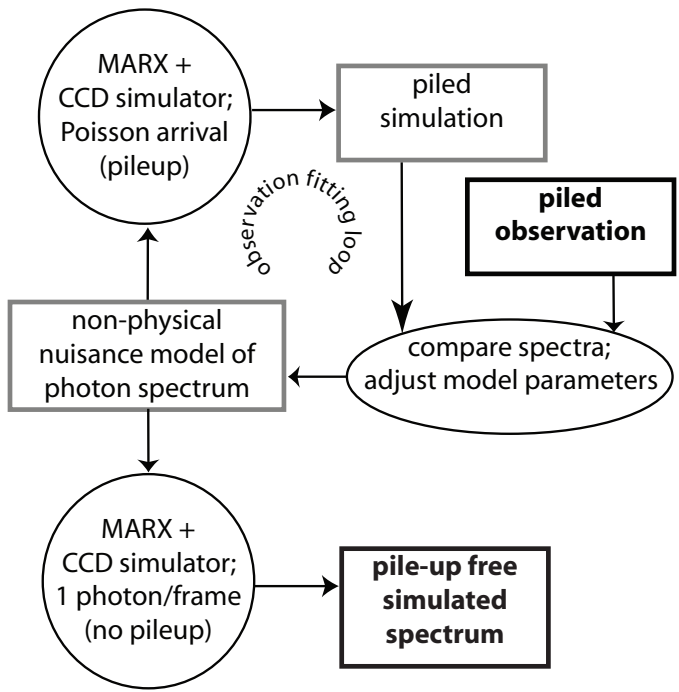} 
\caption{Data flow diagram for reconstructing a pile-up-free \ACIS\ spectrum (bottom right) from a piled observed \ACIS\ spectrum (top right).
No interpretation is attempted on the non-physical model of an incident photon spectrum that reproduces the piled observation.
\label{recon_spectrum_diagram.fig}}
\end{figure}

Figure~\ref{recon_spectrum_diagram.fig} outlines a new variant to the forward modeling approach.  Here a non-physical model of the source spectrum with many free parameters feeds an incident photon spectrum to the \anchorfoot{http://space.mit.edu/cxc/marx/}{\MARX\ mirror simulator}, which produces simulated photons by modeling the \Chandra-\ACIS\ PSF and the observatory dither pattern\footnote
{
During an observation, \Chandra\ executes a slow pattern of motion around the nominal target coordinates known as \anchorparen{http://cxc.harvard.edu/ciao/why/dither.html}{dither}.
}.  
A model of the \ACIS\ CCD \citep{Townsley00,Townsley02a,Townsley02b}\footnote
{\citet{Townsley02a} and \citet{Townsley02b} are available in the Physics database of ADS.}
simulates (1) the physical interaction of those photons with the CCD, (2) the process of reading out charge from the CCD (including \anchorfoot{http://cxc.harvard.edu/ciao/why/cti.html}{charge transfer inefficiency} effects), (3) the process of detecting and \anchorfoot{http://cxc.harvard.edu/ciao/dictionary/grade.html}{grading} X-ray events, (4) the event list cleaning process that is applied to actual \ACIS\ data, (5) assignment of event positions on the sky to remove the effects of the observatory's dithered pointing, and (6) the extraction of events within the aperture chosen by the observer.
Since the simulated photons have random arrival times, photon pile-up occurs naturally within the simulation in the form of superposition of electron charge clouds within individual CCD frames.
The non-physical spectral model is iteratively adjusted until the piled simulated spectrum is similar to the observed spectrum, both in shape and in overall count rate.
When an observed piled-up spectrum is reproduced in a satisfactory manner, the simulation is run one final time with pile-up disabled, by allowing only a single photon to arrive in each CCD frame.  
The reconstruction process thus produces two sets of simulated events derived from the same set of incident photons; the first exhibits pile-up due to Poisson arrival of the photons and resembles the observed spectrum, while the second is free from pile-up.

\revise{In principle, an {\em astrophysical} spectral model could be used in this fitting process, and then directly interpreted scientifically, without a need to simulate an \ACIS\ spectrum free from pile-up.
In such a scheme, the simulator would replace the response files that a fitting package like \XSPEC\ uses to model the detector.
However, our CCD simulator is not calibrated to reproduce the response of \ACIS\ perfectly (e.g., time- and spatial variation of the \ACIS\ optical blocking filter transmission, time-varying charge transfer inefficiency, time-varying amplifier gains).
Thus, we choose to fit a non-physical spectral model, then simulate an \ACIS\ spectrum free from pile-up, and then fit that spectrum to astrophysical models in the normal way.
We assume that small imperfections in the calibration of the simulation simply distort the input non-physical spectral model, leaving the pile-up phenomenon (superposition of photon charge clouds) in the simulation largely unperturbed.
Thus, we expect that the simulated pile-up free event list is similar to what \ACIS\ itself would have produced if pile-up were not present.

The observed (piled) spectrum is grouped (e.g., so that each group has a SNR of ${\sim}$5) and the non-physical spectral model is parameterized by an independent photon flux for the energy range covered by each group.
Within each iteration of the fitting process, the difference between the simulated and observed spectrum in each group is used to revise the corresponding photon flux parameter.
This process is analogous to Lucy-Richardson image reconstruction; the redistribution of X-ray photon energy to observed event energy corresponds to the PSF of a telescope blurring the sky onto an observed image.
The very flexible, non-physical model of the input photon spectrum we employ corresponds to the very flexible, non-physical model of the sky employed by Lucy-Richardson image reconstruction, which involves a free parameter for every pixel in the image.
In both situations, the resolution of the model (size of the spectral group, or image pixel size) is usually smaller than the resolution of the instrument (energy resolution of \ACIS, or PSF of a telescope).
In both cases, Poisson noise will add features to the observation (spectrum or image) that are narrower than the energy resolution of the instrument, and thus cannot be reproduced by the model.
This noise will tend to be amplified in the model (appearing as ``salt-and-pepper'' pixels in reconstructed images), as it strives to reproduce the noisy observation.
However, the noisy photon flux model we derive is not interpreted in any way.  
It merely feeds a final pile-up free simulation, in which the noise in the model cannot produce features in the simulated spectrum that are any sharper or larger than the Poisson noise in the observation that sculpted the model.
}

The Davis model makes no use of any explicit information about the aperture that was used to extract X-ray events from the observation of a point source; indeed, that model can be used with diffuse sources.
The original description of the model \citep{Davis01} suggests a very generous point source aperture, and we are not aware of other documents that discuss the use of non-standard apertures.
One of the attractive features of the pile-up reconstruction method outlined here is significant flexibility when choosing an extraction aperture.
For example, this method can be applied to a piled source in a crowded region extracted with a reduced aperture to avoid a neighbor, as is done for two sources in Table~\ref{pile-up_risk.tbl}.   
A non-standard aperture can be particularly helpful when a source suffers from severe pile-up,  when the correlation between observed event rate and incident photon rate weakens or even becomes negative.\footnote
{
See Figure~3 in the \anchorparen{http://cxc.harvard.edu/ciao/download/doc/pileup_abc.pdf}{The \Chandra\ ABC Guide to Pileup}.
}
Use of an annular aperture in such a case discards the severely piled core of the PSF, restoring a positive correlation between observed event rate and incident photon rate, which this method requires in order to derive a model for the incident photon spectrum.

\subsection{Application to Carina Sources}

Detailed characterization of the performance of this spectral reconstruction technique will be performed in future studies.
Ideally, analyses of piled observations would be judged against pile-up-free reference observations, using more appropriate targets that do not suffer the time variability expected for the massive stars exhibiting pile-up in the CCCP.
In this study, we can roughly judge the reasonableness of the spectral reconstruction method by comparing its results to those obtained by the standard pile-up analysis technique, the ``Davis model'' described above.

For this comparison we choose the two most piled Carina sources from Table~\ref{pile-up_risk.tbl} that have the large extraction apertures required for the Davis model: WR~25 and QZ~Car.
We assess the consistency between the methods by simultaneously fitting the piled and reconstructed spectra in \XSPEC\ \citep{Arnaud96} to a single astrophysical model, and then examining the residuals of the fit.
The \XSPEC\ model for the piled spectrum includes a Davis model component to account for pile-up.\footnote
{
More precisely, \XSPEC\ compares each of the two spectra to separate instances of a Davis+astrophysical model.
The astrophysical parameters of the two model instances are tied together.
The Davis model component for the reconstructed spectrum is effectively disabled by setting the frame time parameter to a very small value, and freezing all model parameters.
The independent Davis model component for the piled spectrum allows the {\em alpha} and {\em psffrac} parameters to vary, and sets the frame time parameter to the value that appropriately accounts for the ``FRACEXPO'' effect, as described in the \anchorparen{http://cxc.harvard.edu/ciao/download/doc/pileup_abc.pdf}{The \Chandra\ ABC Guide to Pileup}.
}

\begin{figure}[htb]
\centering
\includegraphics[width=0.6\textwidth]{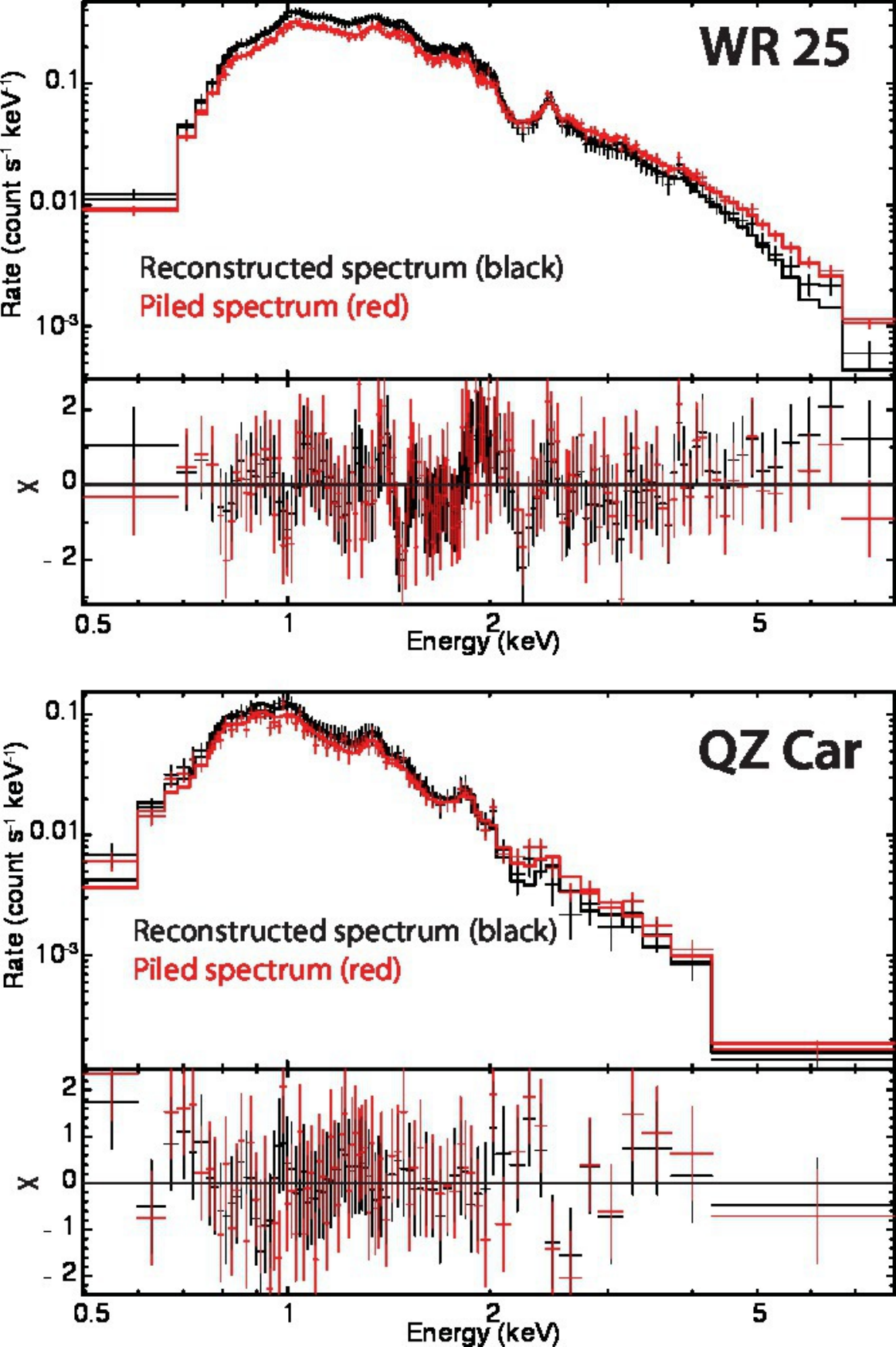} 
\caption{Observed (red) piled spectra and pile-up corrected (black) spectra of the Carina stars WR~25 (top) and QZ~Car (bottom).  
The piled and corrected spectra are fit to a common astrophysical model, using the Davis model to account for pile-up in the former.
Observed spectra are shown as data points with error bars in the upper sub-panels; models are shown as histograms; fit residuals are shown in the lower sub-panels.
\label{recon+Davis.fig}
} 
\end{figure}

Figure~\ref{recon+Davis.fig} shows the resulting piled (red) and pile-up corrected (black) spectra, best fit models, and fit residuals. 
The standard effects of pile-up are seen in the relationship between the corrected and piled spectra: pile-up has reduced the overall count rate, but has hardened the spectrum by creating spurious events at high energy by combining two or more low energy photons.
The most important aspect of these spectra, however, are their model residuals.
The magnitude of the residuals for the reconstructed spectra and their similarity to the residuals for the piled spectra demonstrate the fidelity of the method introduced here, and provide additional confirmation of the Davis model.

\acknowledgements Acknowledgments:  
We appreciate the time and useful suggestions contributed by our anonymous referee.
This work is supported by Chandra X-ray Observatory grant GO8-9131X (PI:  L.\ Townsley) and by the ACIS Instrument Team contract SV4-74018 (PI:  G.\ Garmire), issued by the {\em Chandra} X-ray Center, which is operated by the Smithsonian Astrophysical Observatory for and on behalf of NASA under contract NAS8-03060.
This publication makes use of data products from the Two Micron All Sky Survey, which is a joint project of the University of Massachusetts and the Infrared Processing and Analysis Center/California Institute of Technology, funded by the National Aeronautics and Space Administration and the National Science Foundation.
This work is based in part on observations made with the {\em Spitzer Space Telescope}, which is operated by the Jet Propulsion Laboratory, California Institute of Technology under a contract with NASA.

{\em Facilities:} \facility{CXO (ACIS)}, \facility{Spitzer (IRAC)}, \facility{VLT:Yepun (HAWK-I)}, \facility{CTIO:2MASS ()}


\end{document}

%% file: xray_column_labels.tex
\begin{deluxetable}{lll}
\tablecaption{CCCP X-ray Sources and Properties \label{xray_properties.tbl}
}
\tablecolumns{3}
\tablewidth{0pt}
\tabletypesize{\scriptsize}

\tablehead{
\colhead{Column Label} & \colhead{Units} & \colhead{Description} \\
}
\startdata
  Name                       & \nodata              & \parbox[t]{3in}{IAU source name; prefix is CXOGNC~J \\({\em Chandra X-ray Observatory} Great Nebula in Carina)}\\
  Label\tnm{a}               & \nodata              & source name used within the CCCP project       \\
  RAdeg                      & deg                  & right ascension (J2000)                                         \\
  DEdeg                      & deg                  & declination (J2000)                                             \\
  PosErr                     & arcsec               & 1-$\sigma$ error circle around (RAdeg,DEdeg)                               \\
  PosType                    & \nodata              & algorithm used to estimate position \citep[][\S7.1]{Broos10}  \\
                            &                      &                                                                 \\
  ProbNoSrc\_min             & \nodata              & smallest of ProbNoSrc\_t, ProbNoSrc\_s, ProbNoSrc\_h      \\
  ProbNoSrc\_t               & \nodata              & p-value\tnm{b} for no-source hypothesis \citep[][\S4.3]{Broos10} \\
  ProbNoSrc\_s               & \nodata              & p-value for no-source hypothesis                                \\
  ProbNoSrc\_h               & \nodata              & p-value for no-source hypothesis                                \\
                            &                      &                                                                 \\
  ProbKS\_single\tnm{c}      & \nodata              & \parbox[t]{3in}{smallest p-value for the one-sample Kolmogorov-Smirnov statistic under the no-variability null hypothesis within a single-observation}\\
  ProbKS\_merge\tnm{c}       & \nodata              & \parbox[t]{3in}{smallest p-value for the one-sample Kolmogorov-Smirnov statistic under the no-variability null hypothesis over merged observations}      \\
                            &                      &                                                                 \\
  ExposureTimeNominal        & s                    & total exposure time in merged observations                     \\
  ExposureFraction\tnm{d}    & \nodata              & fraction of ExposureTimeNominal that source was observed\\
  NumObservations            & \nodata              & total number of observations extracted                                \\
  NumMerged                  & \nodata              & \parbox[t]{3in}{number of observations merged to estimate photometry properties}\\
  MergeBias                  & \nodata              & fraction of exposure discarded in merge\\
                            &                      &                                                                 \\
  Theta\_Lo                  & arcmin               & smallest off-axis angle for merged observations                 \\
  Theta                      & arcmin               &  average off-axis angle for merged observations                 \\
  Theta\_Hi                  & arcmin               &  largest off-axis angle for merged observations                 \\
                            &                      &                                                                 \\
  PsfFraction                & \nodata              & average PSF fraction (at 1.5 keV) for merged observations \\
  SrcArea                    & (0.492 arcsec)$^2$   & average aperture area for merged observations                   \\
  AfterglowFraction\tnm{e}   & \nodata              & suspected afterglow fraction                   \\
                            &                      &                                                                 \\
  SrcCounts\_t               & count                & observed counts in merged apertures                             \\
  SrcCounts\_s               & count                & observed counts in merged apertures                             \\
  SrcCounts\_h               & count                & observed counts in merged apertures                             \\
                            &                      &                                                                 \\
  BkgScaling                 & \nodata              & scaling of the background extraction \citep[][\S5.4]{Broos10}                 \\
                            &                      &                                                                 \\
  BkgCounts\_t               & count                & observed counts in merged background regions                    \\
  BkgCounts\_s               & count                & observed counts in merged background regions                    \\
  BkgCounts\_h               & count                & observed counts in merged background regions                    \\
                            &                      &                                                                 \\
  NetCounts\_t               & count                & net counts in merged apertures                                  \\
  NetCounts\_s               & count                & net counts in merged apertures                                  \\
  NetCounts\_h               & count                & net counts in merged apertures                                  \\
                            &                      &                                                                 \\
  MeanEffectiveArea\_t\tnm{f}& cm$^2$~count~photon$^{-1}$ & mean ARF value                                                  \\
  MeanEffectiveArea\_s       & cm$^2$~count~photon$^{-1}$ & mean ARF value                                                  \\
  MeanEffectiveArea\_h       & cm$^2$~count~photon$^{-1}$ & mean ARF value                                                  \\
                            &                      &                                                                 \\
  MedianEnergy\_t\tnm{g}     & keV                  & median energy, observed spectrum                                \\
  MedianEnergy\_s            & keV                  & median energy, observed spectrum                                \\
  MedianEnergy\_h            & keV                  & median energy, observed spectrum                                \\
                            &                      &                                                                 \\
  EnergyFlux\_t              & erg~cm$^{-2}$~s$^{-1}$ & max(EnergyFlux\_s,0) + max(EnergyFlux\_h,0)                               \\
  EnergyFlux\_s              & erg~cm$^{-2}$~s$^{-1}$ & See Eqn.~\ref{EnergyFlux.eqn}. \\ 
  EnergyFlux\_h              & erg~cm$^{-2}$~s$^{-1}$ & See Eqn.~\ref{EnergyFlux.eqn}. \\ 
                            &                      &                                                                 \\
  NetCounts\_Lo\_t\tnm{h}    & count                & 1-sigma lower bound on NetCounts\_t               \\
  NetCounts\_Hi\_t           & count                & 1-sigma upper bound on NetCounts\_t               \\
  \\
  NetCounts\_Lo\_s           & count                & 1-sigma lower bound on NetCounts\_s               \\
  NetCounts\_Hi\_s           & count                & 1-sigma upper bound on NetCounts\_s               \\
  \\
  NetCounts\_Lo\_h           & count                & 1-sigma lower bound on NetCounts\_h               \\
  NetCounts\_Hi\_h           & count                & 1-sigma upper bound on NetCounts\_h               
\enddata
                                                                                                                                              
\tablecomments{
Column labels are shown as they appear in Vizier, where formatted labels are not possible.
}
\tablecomments{
The first 14,368 sources---the catalog used in other CCCP studies---are sorted by RA; the 14,369$^{th}$ source was identified too late in the data analysis process to be conveniently inserted at its rightful position in the table.
}
\tablecomments{
The suffixes ``\_t'', ``\_s'', and ``\_h'' on names of photometric quantities designate the {\em total} (0.5--8~keV), {\em soft} (0.5--2~keV), and {\em hard} (2--8~keV) energy bands. 
}
\tablecomments{
Source significance quantities (ProbNoSrc\_t, ProbNoSrc\_s, ProbNoSrc\_h, ProbNoSrc\_min) are computed using a subset of each source's extractions chosen to maximize significance \citep[][\S6.2]{Broos10}.
Source position quantities (RAdeg, DEdeg, PosErr) are computed using a subset of each source's extractions chosen to minimize the position uncertainty \citep[][\S6.2 and 7.1]{Broos10}.   
All other quantities are computed using a subset of each source's extractions chosen to balance the conflicting goals of minimizing photometric uncertainty and of avoiding photometric bias \citep[][\S6.2 and 7]{Broos10}. 
}

\tablenotetext{a}{Source labels identify a CCCP pointing \citep[][Table~1]{Townsley11a}; they do not convey membership in astrophysical clusters.}

\tablenotetext{b}{In statistical hypothesis testing, the p-value is the probability of obtaining a test statistic at least as extreme as the one that was actually observed when the null hypothesis is true.}

\tablenotetext{c}{See \citet[][\S7.6]{Broos10} for a description of the variability metrics, and caveats regarding possible spurious indications of variability using the ProbKS\_merge metric. }

\tablenotetext{d}{Due to dithering over inactive portions of the focal plane, a \Chandra\ source is often not observed during some fraction of the nominal exposure time.  (See \url{http://cxc.harvard.edu/ciao/why/dither.html}.)  The reported quantity is FRACEXPO produced by the \CIAO\ tool {\em mkarf}.}

\tablenotetext{e}{Since the extracted event data are lightly-cleaned to avoid removing legitimate X-ray events from bright sources \citep[][\S3]{Broos10}, some background events arising from an effect known as ``afterglow'' (\url{http://cxc.harvard.edu/ciao/why/afterglow.html}) will remain.
We attempt to identify afterglow events using the tool \url{ae_afterglow_report} (see the \AEacro\ manual), and report the fraction of extracted events attributed to afterglow.}

\tablenotetext{f}{The ancillary response file (ARF) in \ACIS\ data analysis represent both the effective area of the observatory and the fraction of the observation for which data were actually collected for the source (ExposureFraction).}

\tablenotetext{g}{MedianEnergy is the \AEacro\ quantity ENERG\_PCT50\_OBSERVED, the median energy of extracted events, corrected for background \citep[][\S7.3]{Broos10}. }

\tablenotetext{h}{Confidence intervals (68\%) for NetCounts quantities are estimated by the \CIAO\ tool {\em aprates} (\url{http://asc.harvard.edu/ciao/ahelp/aprates.html}).}

\end{deluxetable}

%% file: counterpart_column_labels.tex
\begin{deluxetable}{rll}                        
\tablecaption{Counterpart Properties \label{counterpart_properties.tbl}
}
\tablecolumns{3}
\tablewidth{0pt}
\tabletypesize{\footnotesize}

\tablehead{
\colhead{Column Label} & \colhead{Units} & \colhead{Description} 
}
\startdata
  Name                 & \nodata              & official name of X-ray source (prefix is CXOGNC J)    \\
  Label                & \nodata              & X-ray source name used within the CCCP project \\
  
\tableline  
  Offset(Skiff)        & arcsec              & offset from matching ACIS source       \\
  Offset(KR)           & arcsec              & offset from matching ACIS source       \\
  Offset(PPMXL)        & arcsec              & offset from matching ACIS source       \\
  Offset(UCAC3)        & arcsec              & offset from matching ACIS source       \\
  Offset(BSS)          & arcsec              & offset from matching ACIS source       \\
  Offset(CMD)          & arcsec              & offset from matching ACIS source       \\
  Offset(DETWC)        & arcsec              & offset from matching ACIS source       \\
  Offset(MDW)          & arcsec              & offset from matching ACIS source       \\
  Offset(MJ)           & arcsec              & offset from matching ACIS source       \\
  Offset(CP)           & arcsec              & offset from matching ACIS source       \\
  Offset(DAY)          & arcsec              & offset from matching ACIS source       \\
  Offset(HAWK-I)       & arcsec              & offset from matching ACIS source       \\
  Offset(2MASS)        & arcsec              & offset from matching ACIS source       \\
  Offset(SOFI)         & arcsec              & offset from matching ACIS source       \\
  Offset(NACO)         & arcsec              & offset from matching ACIS source       \\
  Offset(Sana)         & arcsec              & offset from matching ACIS source       \\
  Offset(SpVela)       & arcsec              & offset from matching ACIS source       \\
  Offset(SpSmith)      & arcsec              & offset from matching ACIS source       \\
  Offset(AC)           & arcsec              & offset from matching ACIS source       \\
  
\tableline
Identifier             & \nodata              & historical star name          \\
SpType                 & \nodata              & spectral type                                           \\
SpRef                  & \nodata              & reference for spectral type \\


\tableline  
  Id(KR)               & \nodata              & row identifier ({\em recno} in published catalog)         \\
Bmag(KR)               & mag                  & B band magnitude                       \\
Vmag(KR)               & mag                  & V band magnitude                       \\

\tableline  
  Id(PPMXL)            & \nodata              & row identifier ({\em ipix} in published catalog)         \\
  
\tableline  
  Id(UCAC3)            & \nodata              & row identifier ({\em 3UC} in published catalog)         \\
  
\tableline  
  Id(BSS)              & \nodata              & row identifier ({\em ID} in published catalog)       \\

\tableline  
  Id(CMD)              & \nodata              & row identifier ({\em recno} in published catalog)         \\
Vmag(CMD)              & mag                  & V band magnitude                               \\
 B-V(CMD)              & mag                  & B-V color                                      \\
 
\tableline  
  Id(DETWC)            & \nodata              & row identifier ({\em RAJ2000+DEJ2000} in published catalog) \\
Vmag(DETWC)            & mag                  & V band magnitude                       \\
 U-B(DETWC)            & mag                  & U-B color                              \\
 B-V(DETWC)            & mag                  & B-V color                              \\
 
\tableline  
  Id(MDW)              & \nodata              & row identifier ({\em Name} in published catalog)         \\
 
\tableline  
  Id(MJ)               & \nodata              & row identifier ({\em Star} in published catalog) \\
Vmag(MJ)               & mag                  & V band magnitude                       \\
 U-B(MJ)               & mag                  & U-B color                              \\
 B-V(MJ)               & mag                  & B-V color                              \\
 
\tableline  
  Id(CP)               & \nodata              & row identifier ({\em ID} in published catalog) \\
Umag(CP)               & mag                  & U band magnitude                       \\
Bmag(CP)               & mag                  & B band magnitude                       \\
Vmag(CP)               & mag                  & V band magnitude                       \\
Rmag(CP)               & mag                  & R band magnitude                       \\
Imag(CP)               & mag                  & I band magnitude                       \\

\tableline  
  Id(DAY)              & \nodata              & row identifier ({\em Name} in published catalog) \\
Vmag(DAY)              & mag                  & Johnson V magnitude                    \\
 U-B(DAY)              & mag                  & Johnson U-B colour index               \\
 B-V(DAY)              & mag                  & Johnson B-V colour index               \\
V-Rc(DAY)              & mag                  & Johnson-Cousins V-Rc colour index      \\
V-Ic(DAY)              & mag                  & Johnson-Cousins V-Ic colour index      \\
  AV(DAY)              & \nodata              & Absorption in V band                   \\
  
\tableline  
  Id(HAWK-I)           & \nodata              & row identifier ({\em RA+DEC} in published catalog)        \\
  
\tableline  
      Id(2MASS)         & \nodata              & row identifier ({\em Designation} in published catalog)  \\
    J\_M(2MASS)         & mag                  &  J-band magnitude                                  \\
    H\_M(2MASS)         & mag                  &  H-band magnitude                                  \\
    K\_M(2MASS)         & mag                  & Ks-band magnitude                                  \\
Ph\_Qual(2MASS)         & \nodata              & Photometric quality flag                           \\
 CC\_Flg(2MASS)         & \nodata              & Contamination and confusion flag                   \\

\tableline  
   Id(SOFI)            & \nodata              & row identifier ({\em recno} in published catalog)         \\
 Jmag(SOFI)            & mag                  &  J-band magnitude                                   \\
 Hmag(SOFI)            & mag                  &  H-band magnitude                                   \\
Ksmag(SOFI)            & mag                  & Ks-band magnitude                                   \\

\tableline  
   Id(NACO)            & \nodata              & row identifier ({\em recno} in published catalog)         \\
 Jmag(NACO)            & mag                  &  J-band magnitude                                   \\
 Hmag(NACO)            & mag                  &  H-band magnitude                                   \\
Ksmag(NACO)            & mag                  & Ks-band magnitude                                   \\

\tableline  
  Id(Sana)             & \nodata              & row identifier ({\em recno} in published catalog)         \\
 Hmag(Sana)            & mag                  &  H-band magnitude                                   \\
Ksmag(Sana)            & mag                  & Ks-band magnitude                                   \\

\tableline  
  Id(SpVela)           & \nodata              & row identifier ({\em DESIG} in published catalog)       \\
  mag3\_6(SpVela)      & mag                  & 3.6 \micron\ magnitude                                 \\
  mag4\_5(SpVela)      & mag                  & 4.5 \micron\ magnitude                                 \\
  mag5\_8(SpVela)      & mag                  & 5.8 \micron\ magnitude                                 \\
  mag8\_0(SpVela)      & mag                  & 8.0 \micron\ magnitude                                 \\
  
\tableline  
  Id(SpSmith)          & \nodata              & row identifier ({\em DESIG} in published catalog)      \\
  mag3\_6(SpSmith)     & mag                  & 3.6 \micron\ magnitude                                 \\
  mag4\_5(SpSmith)     & mag                  & 4.5 \micron\ magnitude                                 \\
  mag5\_8(SpSmith)     & mag                  & 5.8 \micron\ magnitude                                 \\
  mag8\_0(SpSmith)     & mag                  & 8.0 \micron\ magnitude                                 \\

\tableline  
  Id(AC)               & \nodata              & row identifier ({\em CXOTr16} in published catalog)         

\enddata
\end{deluxetable}